\documentclass[aps,pre,twocolumn,longbibliography,preprintnumbers,10pt]{revtex4-2} 

\usepackage{amsmath} 
\usepackage{amsfonts}
\usepackage{amssymb}
\usepackage{graphicx}
\usepackage{color}
\usepackage{accents}
\usepackage{enumitem}
\usepackage{bbold}
\usepackage{amsmath,amsfonts,amsthm} % Math packages
\usepackage{physics}

\providecommand{\innerprod}[2]{\ensuremath{ \left\langle #1 | #2 \right\rangle}}
\providecommand{\outerprod}[2]{\ensuremath{| #1 \rangle \langle #2 | }}  
\providecommand{\ID}{\ensuremath{\mathbb{1}}} 

\usepackage[colorlinks=true, citecolor=cyan]{hyperref} 
\newcommand{\figone}[0]{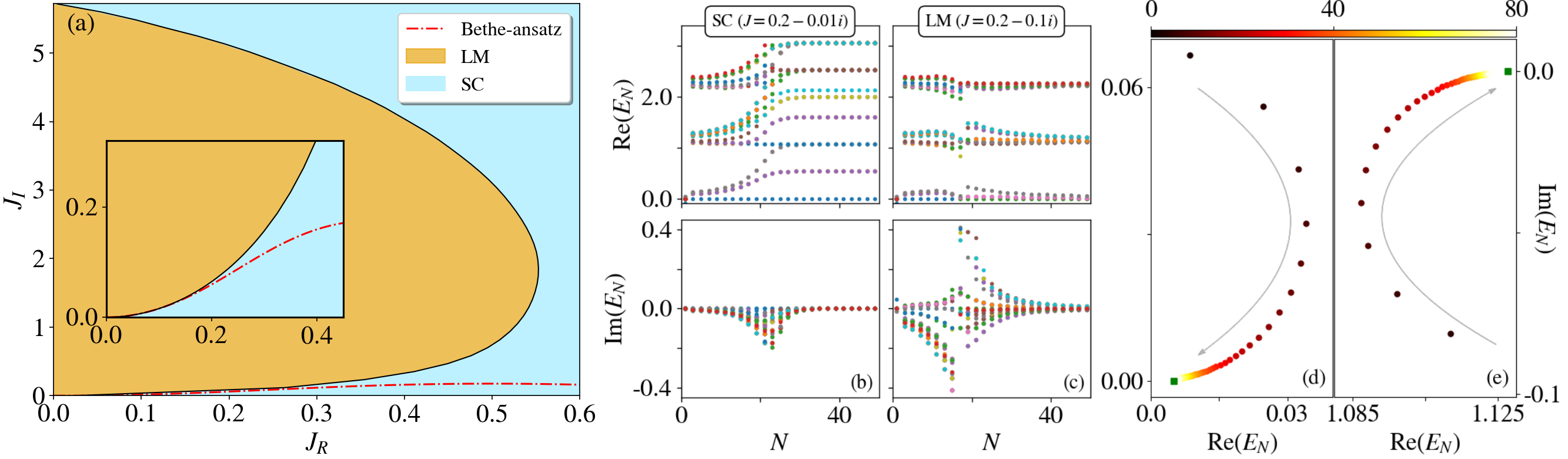}
\newcommand{\figtwo}[0]{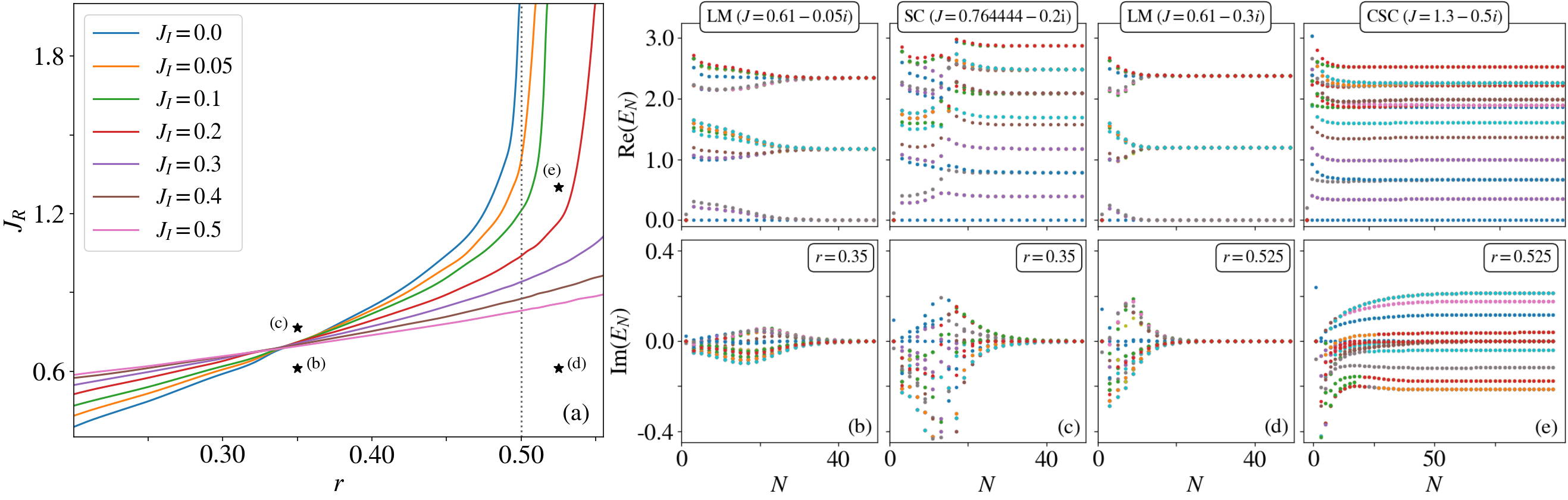}
\newcommand{\figthree}[0]{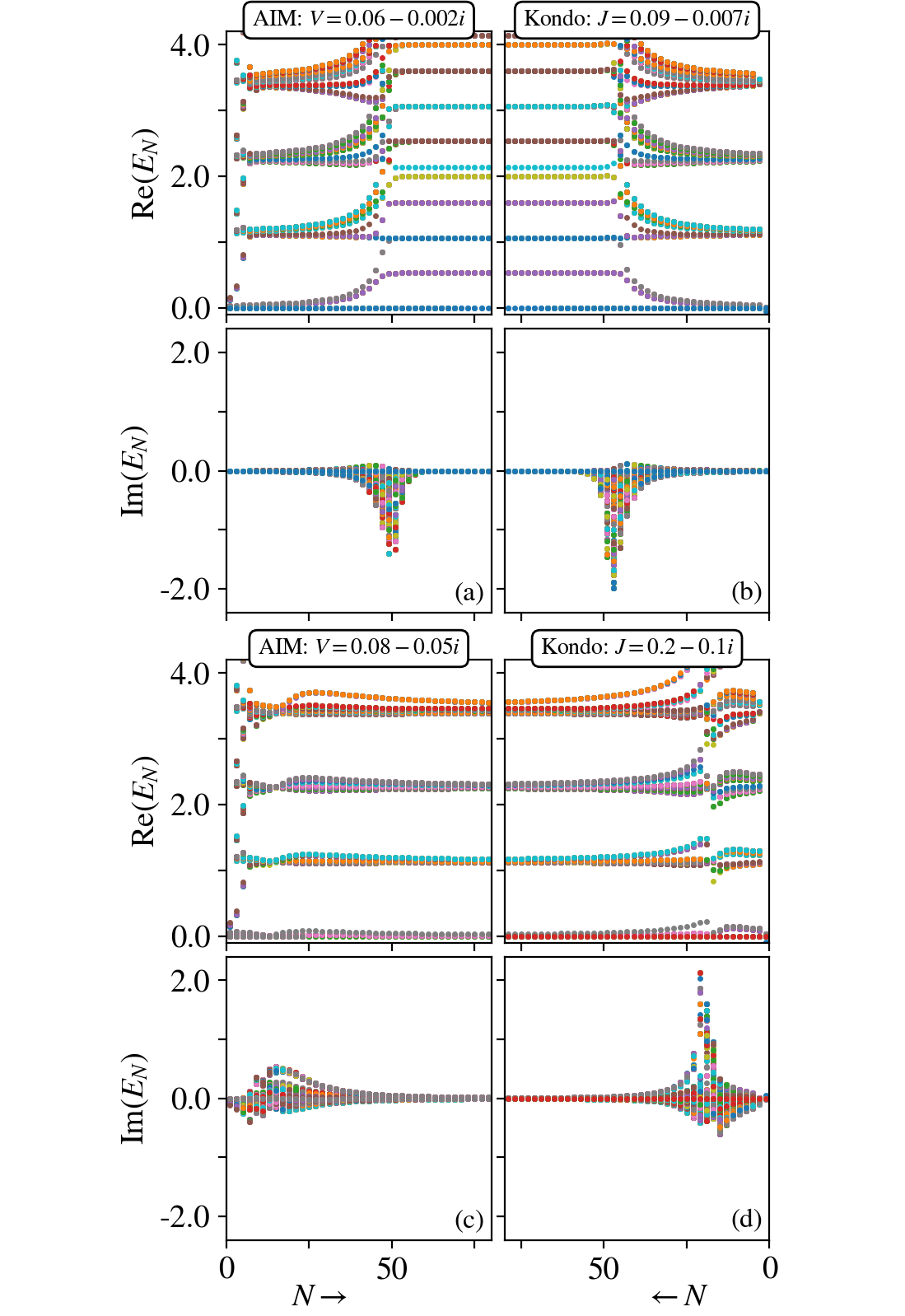}
\newcommand{\figfour}[0]{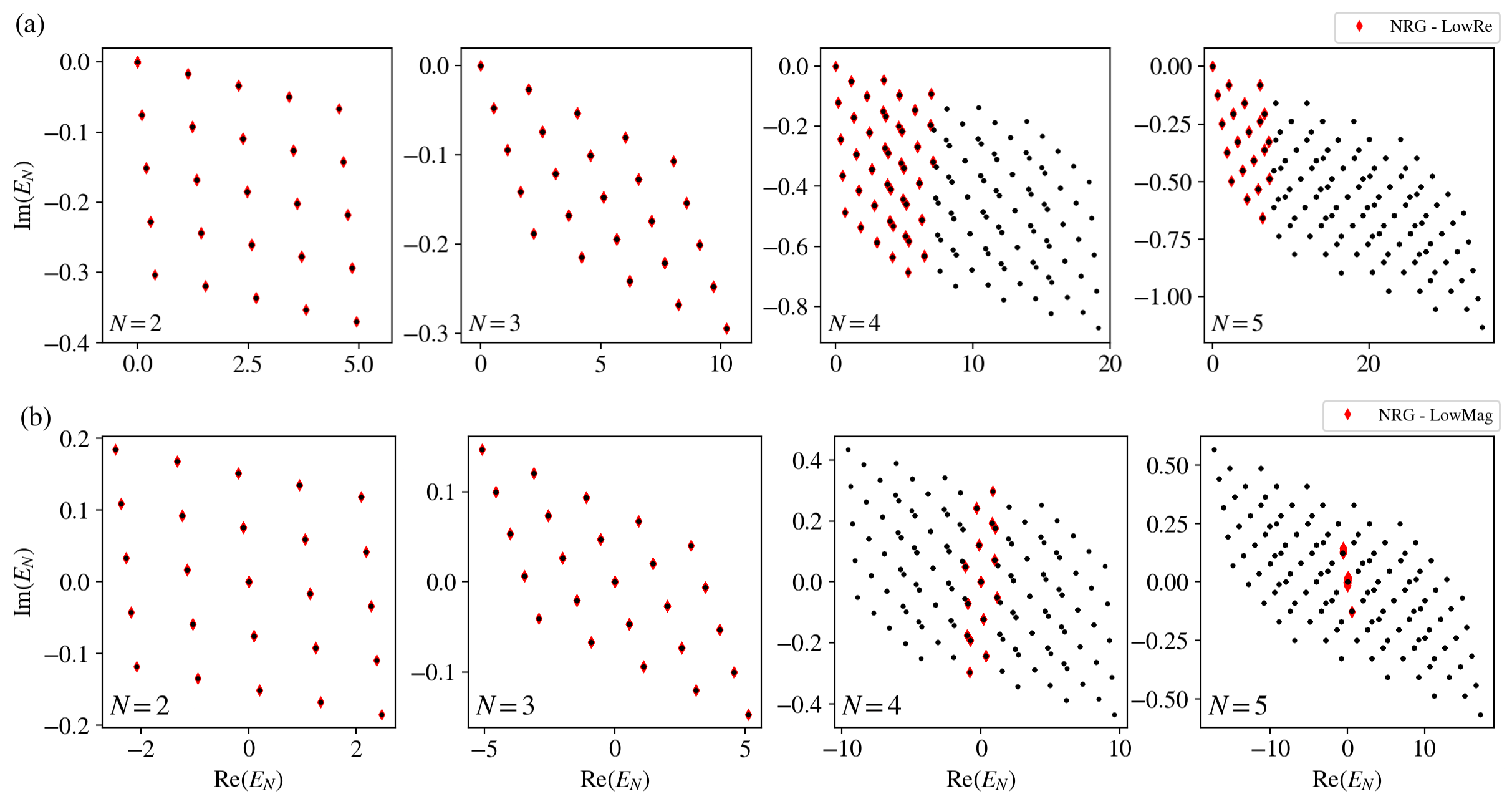}

\begin{document}

\title{Non-Hermitian Numerical Renormalization Group:\\ Solution of the non-Hermitian Kondo model}

\author{Phillip C. Burke}
\affiliation{School of Physics, University College Dublin, Belfield, Dublin 4, Ireland}
\affiliation{Centre for Quantum Engineering, Science, and Technology, University College Dublin, Dublin 4, Ireland}

\author{Andrew K. Mitchell} 
\affiliation{School of Physics, University College Dublin, Belfield, Dublin 4, Ireland}
\affiliation{Centre for Quantum Engineering, Science, and Technology, University College Dublin, Dublin 4, Ireland}

\begin{abstract}
\noindent Non-Hermitian (NH) Hamiltonians describe open quantum systems, nonequilibrium dynamics, and dissipative processes. Although a rich range of single-particle NH physics has been uncovered, many-body phenomena in strongly correlated NH systems have been far less well studied. 
The Kondo effect, an important paradigm for strong correlation physics, has recently been considered in the NH setting. Here we develop a NH generalization of the numerical renormalization group (NRG) and use it to solve the NH Kondo model. Our non-perturbative solution applies beyond weak coupling, and we uncover a nontrivial phase diagram. 
The method is showcased by application to the NH pseudogap Kondo model, which we show supports a completely novel phase with a genuine NH stable fixed point and complex eigenspectrum. Our NH-NRG code, which can be used in regimes and for models inaccessible to, e.g., perturbative scaling and Bethe ansatz, is provided open source.
\end{abstract} 

\maketitle

\graphicspath{{./Figures/}}
%%%%%%%%%%%%%%%%%%%%%%%%%%%%%%%%%%%%%%%%%%%%%%%%%%%%%%%%%%%%%%%%%%%%%%%%%%%%%%%%%%%%%%%%%%

The past two decades have seen immense interest in open quantum systems, with non-Hermitian (NH) Hamiltonians describing the effective dynamics of dissipative systems playing a key role~\cite{Rotter_nonHermitian_JoPA2009, MingantiNori_NonHermEP_PRA2019, Ashida_NonHermPhys_AiP2020, Bergholtz_NonHermTop_RevMod2021, Roccati_NonHermPhys_OSID2022}. NH Hamiltonians present certain unique challenges, such as dealing with complex eigenvalues, non-orthonormal eigenvectors \cite{Brody_BiorthogonalQM_JoPA2014, Wiersig_Nonorthogonality_PRA2018}, and exceptional points \cite{heiss_avoidedEPs_JoPA1990, heiss_chiralityofEPs_EPJD2001, Berry_EPs_CJoP2004, Muller_EPs_JoPA2008, heiss_physicsofEPs_JoPA2012, BurkeHaque_NonHermTB_PRA2020, Sayyad_NonDefectiveEPs_SciPost2023, LiJiang_HigherOrderEPs_PRR2024, Kunst_HigherOrderEPs_PRR2024, ErdoganHaque_HigherOrderEPs_arxiv2024, Gohsrich_HigherOrderEPs_arxiv2024} -- singularities in parameter space at which eigenvalues and eigenstates coalesce. NH systems with $\mathcal{PT}$-symmetry~\cite{Bender_PTSymmetry_PRL1998, Bender_ComplexQM_PRL2002, Bender_nonHermHam_RoPiP2007} are somewhat simpler, having real eigenvalues; but many systems of interest do not fall into this class.  
Much attention has, to date, focused on single-particle NH systems~\cite{Ashida_NonHermPhys_AiP2020, Bergholtz_NonHermTop_RevMod2021}, while many-body counterparts remain far less well explored. Although recent work has begun to address strongly-correlated NH physics, non-perturbative numerical methods beyond exact diagonalization remain limited \cite{Zhong_NHDMRG_arxiv2024, Fang_NonHermDMRG_PRB2025}. 

The Kondo model~\cite{Hewson1993} is a classic paradigm for strong-correlation physics in the standard Hermitian scenario, so the solution of its NH generalizations is naturally of importance for understanding NH physics in the many-body context. Furthermore, as shown in Ref.~\cite{NakagawaKawakamiUeda_NonHermKondo_PRL2018},  ultracold atom systems undergoing inelastic scattering with two-body losses can be described by an effective NH Kondo model. These factors have stimulated considerable interest in a range of NH quantum impurity models~\cite{NakagawaKawakamiUeda_NonHermKondo_PRL2018, Hasegawa_NonHermKondoQuDot_arxiv2021, HanSchultzKim_NHKondoLuttingerLiquid_PRB2023, kattel2024dissipationdrivenphasetransition, chen2024criticalbehaviornonhermitiankondo, Lourenco_PTKondo_PRB2018, Kulkarni_PTAndersonModel_PRB2022, yamamoto2024correlationversusdissipationnonhermitian,kulkarnianderson, Yoshimura_NonHermImpurity_PRB2020, vanhoecke2024kondozenocrossoverdynamicsmonitored, stefanini2024dissipativerealizationkondomodels,
QuStefanini_VariationalDissipativeImpurity_PRB2025}.

The non-Hermitian Kondo model reads,
\begin{eqnarray}
    \label{eq:NHK}
    \hat{H}=\hat{H}_{\rm bath} + J \hat{\boldsymbol{S}}_i\cdot \hat{\boldsymbol{S}}_0 \;,
\end{eqnarray}
where $J=J_R-iJ_I$ is taken to be complex, $\hat{\boldsymbol{S}}_i$ is a spin-$\tfrac{1}{2}$ operator for the impurity, $\hat{H}_{\rm bath}=\sum_{k,\sigma}\epsilon_k c_{k\sigma}^{\dagger}c_{k\sigma}^{\phantom{\dagger}}$ describes a continuum bath of non-interacting conduction electrons labeled by spin $\sigma=\uparrow,\downarrow$ 
 and momentum $k$, and $\hat{\boldsymbol{S}}_0=\tfrac{1}{2}\sum_{\sigma,\sigma'}c_{0\sigma}^{\dagger}\boldsymbol{\tau}_{\sigma,\sigma'} c_{0\sigma'}^{\phantom{\dagger}}$ is the local conduction electron spin density at the impurity position (here $c_{0\sigma}=\sum_k \alpha_k c_{k\sigma}$ and $\boldsymbol{\tau}$ is the Pauli vector). The bath is characterized by its density of states (DOS) at the impurity, $\rho(\omega)$. For a standard metallic flat band, we take $\rho(\omega)=\rho_0\Theta(D-|\omega|)$.  Eq.~\eqref{eq:NHK} does not possess $\mathcal{PT}$-symmetry, so generically has a complex eigenspectrum.

%--\FIGURE\---------------------
\begin{figure*}[t!]
	\centering 
    \includegraphics[width=\linewidth]{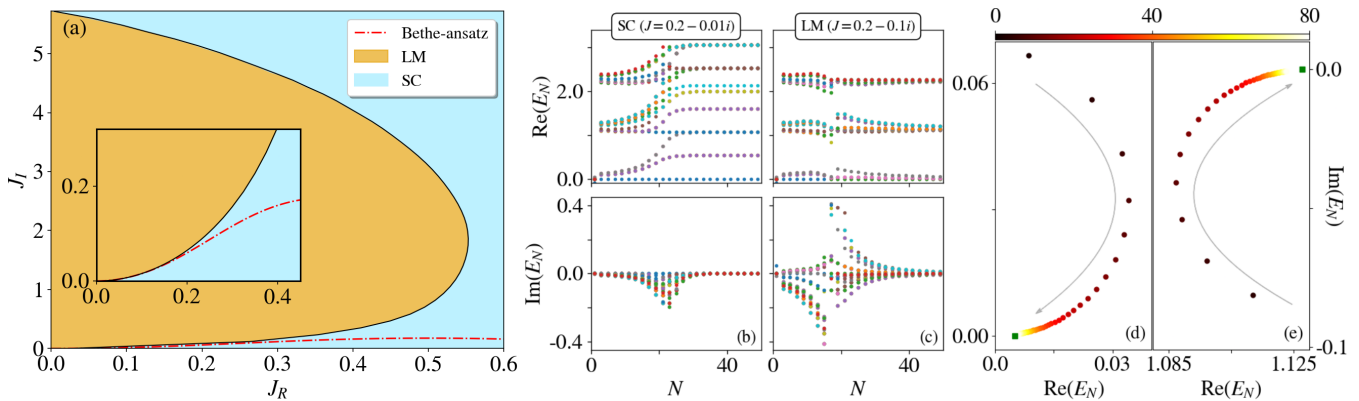}
    \caption{Solution of the non-Hermitian Kondo model using NH-NRG. (a) Phase diagram in the $(J_R,J_I)$ plane, showing the numerically-exact boundary (black line) separating SC (blue) and LM (orange) phases. Red dot-dashed line shows the Bethe ansatz result~\cite{NakagawaKawakamiUeda_NonHermKondo_PRL2018} which is valid for $|J|\lesssim 0.25$ and agrees perfectly with NH-NRG in that regime (see inset). (b,c) RG flow of the NH-NRG complex eigenvalues $E_N$ with iteration number $N$, showing the real and imaginary parts in the top and bottom panels, for representative systems in the SC and LM phases. (d,e) Reversion of eigenvalue RG flow in the Argand plane for an LM system ($J=0.1-0.5i$). $N$ increases in the direction of the arrows towards the Hermitian Kondo fixed point value (green point). Shown for different representative states in (d) and (e). NH-NRG calculations performed for $\Lambda=3$ and $N_k=400$. 
    }
	\label{fig_Kondo_PhaseDiagram}
\end{figure*}
%--\FIGURE\----------------------

The standard Hermitian Kondo model is recovered for $J_I=0$. 
For antiferromagnetic coupling $J_R>0$, the impurity spin is dynamically screened by surrounding conduction electrons via the Kondo effect~\cite{Hewson1993} at low temperatures $T\ll T_K$, with $T_K\sim D e^{-1/\rho_0 J_R}$ the Kondo temperature. The physics is non-perturbative and non-Markovian: even for small bare $J_R$, the impurity becomes strongly coupled to the bath at low $T$ by formation of a many-body Kondo singlet state inside a large entanglement `cloud'~\cite{sorensen1996scaling,mitchell2011real,lee2015macroscopic}. The Kondo effect can be understood in the renormalization group (RG) framework~\cite{anderson1970poor} as a flow from the unstable local moment (LM) fixed point, corresponding to a free spin on the impurity decoupled from the bath, to the stable strong-coupling (SC) fixed point in which the impurity is bound up in the Kondo singlet. A full, non-perturbative solution of the Kondo problem is provided by Wilson's numerical renormalization group (NRG) technique~\cite{wilson1975renormalization,Bulla_NRG_RevMod2008}, which can also be applied to generalized quantum impurity problems, and works with arbitrary bath density of states. The Hermitian Kondo model in the wide flat-band limit can also be solved exactly by Bethe ansatz~\cite{andrei1983solution}.

The NH Kondo model was studied in Ref.~\cite{NakagawaKawakamiUeda_NonHermKondo_PRL2018} using a combination of perturbative scaling and Bethe ansatz, which provides a rather complete picture of the weak-coupling physics up to $|J|/D\lesssim 0.25$, beyond which the methods break down. It was shown that sufficiently strong dissipation (tuned by increasing $J_I$) can produce a quantum phase transition between the standard Kondo SC phase and an unscreened LM phase, via a mechanism analogous to the continuous quantum Zeno effect~\cite{syassen2008strong}. A reversion of the RG flow was observed in the LM phase, which violates the $g$-theorem for Hermitian systems~\cite{affleck1991universal}. The low-energy fixed points were found to be \textit{real}, meaning that the metallic NH Kondo model has an emergent Hermiticity. However, this scenario has recently been challenged, with the alternative Bethe ansatz results of Ref.~\cite{kattel2024dissipationdrivenphasetransition} appearing to show a different phase diagram, with a new phase intervening between SC and LM.

In this Letter, we introduce the non-Hermitian numerical renormalization group (NH-NRG) method, which is fully non-perturbative, and can be applied to a wide range of Kondo or Anderson-type impurity models and their variants. With no restriction on coupling strength, we uncover a nontrivial phase diagram for the NH Kondo model (Fig.~\ref{fig_Kondo_PhaseDiagram}a), showing that at weak-to-moderate coupling, the scenario of Ref.~\cite{NakagawaKawakamiUeda_NonHermKondo_PRL2018} exactly pertains. However, for stronger dissipation (larger values of $J_I$) we find re-entrant Kondo behavior; whereas the LM phase is found to terminate entirely beyond a critical value of $J_R$.  
Unlike the Bethe ansatz and other methods such as conformal field theory that rely on linear dispersion~\cite{andrei1983solution,affleck1991universal}, NH-NRG works with equal ease for any bath DOS. We apply NH-NRG to a NH pseudogap Kondo model, showing that the lower-critical dimension of the Hermitian model~\cite{fritz2004phase} is shifted by finite $J_I$, and an entirely new stable fixed point appears that is fundamentally non-Hermitian.

%%%%%%%%%%%%%%

%--\FIGURE\------------------
\begin{figure*}[t!]
	\centering 
    \includegraphics[width=\linewidth]{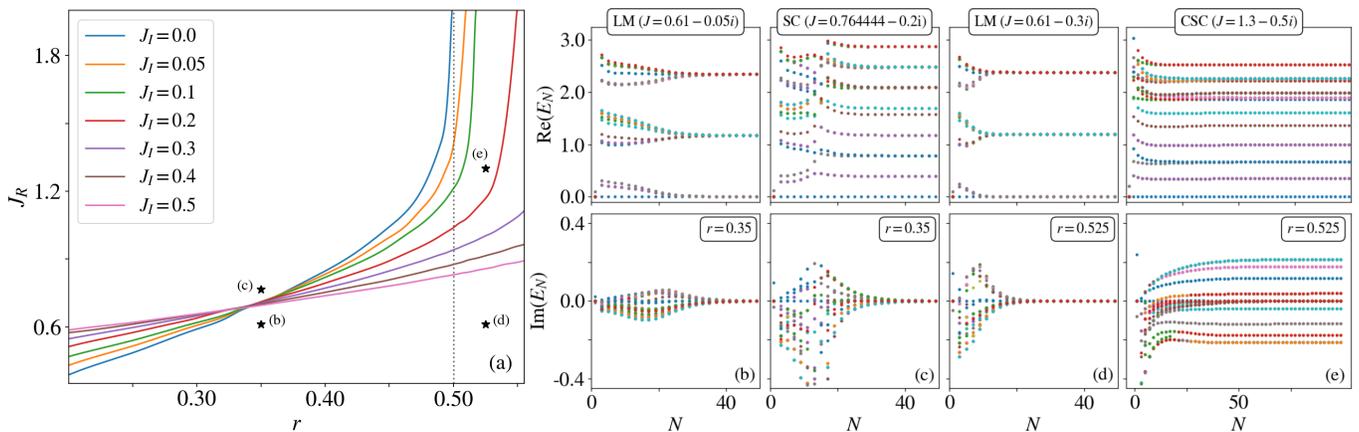}
\caption{Non-Hermitian pseudogap Kondo Model. (a) Critical $J_R$ separating LM and SC phases, vs pseudogap exponent $r$, for different $J_I$. Lower-critical dimension of the Hermitian model at $r=0.5$ shown as the dotted vertical line. (b-e) Eigenvalue RG flow for systems indicated by the star points in (a). (b,c) Representative LM and SC flows for $r=0.35$; (d,e) Flow for LM and a new `complex strong coupling' (CSC) fixed point for $r=0.525$. NH-NRG calculations with $\Lambda=3$ and $N_k=400$.  
    }
	\label{fig_PseudoKondo_PhaseDiagram}
\end{figure*}
%--\FIGURE\----------------

\textit{Non-Hermitian NRG.--}
Here we generalize the standard NRG methodology to treat NH quantum impurity problems. Although the basic algorithm proceeds along similar lines to Wilson's original formulation for Hermitian systems~\cite{wilson1975renormalization,Bulla_NRG_RevMod2008}, incorporating NH physics involves additional challenges. Below we describe the key points, but full technical details and validation checks are given in the \textit{End Matter} and Supplemental Material~\cite{SuppMat}.

In the standard NRG procedure for the Kondo model, the first step is to logarithmically discretize the free conduction electron bath and map it to a 1d Wilson chain (WC). This is done by dividing up the density of states $\rho(\omega)$ into intervals according to the discretization points $\pm D\Lambda^{-n}$, where $\Lambda>1$ is the NRG discretization parameter and $n=0,1,2,3,...$. The continuous electronic density in each interval is replaced by a single pole at the average position with the same total weight, yielding $\rho_{\rm disc}(\omega)$. We then map $\hat{H}_{\rm bath}\to \hat{H}_{\rm WC} = \sum_{n=0}^{\infty}\sum_{\sigma}  (\epsilon_n^{\phantom{\dagger}} f_{n\sigma}^{\dagger} f_{n\sigma}^{\phantom{\dagger}} + t_n^{\phantom{\dagger}}f_{n\sigma}^{\dagger} f_{n+1\sigma}^{\phantom{\dagger}}+ t_nf_{n+1\sigma}^{\dagger} f_{n\sigma}^{\phantom{\dagger}})$ with the real parameters $\{\epsilon_n\}$ and $\{t_n\}$ chosen such that the local density of states at orbital $f_{0\sigma}$ to which the impurity couples is precisely $\rho_{\rm disc}(\omega)$. Due to the logarithmic discretization \cite{wilson1975renormalization}, the WC parameters decay asymptotically as $\sim D\Lambda^{-n/2}$. We now define a sequence of Hamiltonians $\hat{H}_N$ comprising the impurity and the first $N$ chain sites, satisfying the recursion relation $\hat{H}_{N}=\hat{H}_{N-1} +\hat{T}_{N}$ where $\hat{T}_N=\sum_{\sigma}(\epsilon_N^{\phantom{\dagger}} f_{N\sigma}^{\dagger} f_{N\sigma}^{\phantom{\dagger}} + t_{N-1}^{\phantom{\dagger}}f_{N-1\sigma}^{\dagger} f_{N\sigma}^{\phantom{\dagger}}+ t_{N-1}f_{N\sigma}^{\dagger} f_{N-1\sigma}^{\phantom{\dagger}})$. The sequence is initialized by $\hat{H}_{0}=J \hat{\boldsymbol{S}}_i\cdot \hat{\boldsymbol{S}}_0$~\footnote{The spin density of the discretized bath at the impurity position is  $\hat{\boldsymbol{S}}_0=\tfrac{1}{2}\sum_{\sigma,\sigma'}f_{0\sigma}^{\dagger}\boldsymbol{\tau}_{\sigma,\sigma'}f_{0\sigma'}$.} and the full (discretized) model is obtained as $N\to \infty$. Starting from the impurity, we build up the chain by successively adding WC sites using this recursion. At each step $N$, the Hamiltonian $\hat{H}_N$ is diagonalized, and only the $N_k$ lowest energy states are retained to construct the Hamiltonian $\hat{H}_{N+1}$ at the next step. In such a way, we focus on progressively lower energy scales with each iteration. The higher energy states can be discarded at each step due to the energy-scale separation down the chain. 
The RG character of the problem can be seen directly in the evolution with $N$ of the many-particle NRG energy levels of $\hat{H}_N$. This is done by specifying the NRG energies $E_N$ with respect to the ground state energy of that iteration, and then rescaling by a factor $\Lambda^{N/2}$, so that the $N_k$ retained states at each step always span the same energy range. Importantly, the NRG energy levels flow between fixed points (e.g.~from LM to SC). The calculation scales linearly in $N$, and the stable fixed point is reached after a finite number of steps. NRG is thus able to capture an exponentially-wide range of energy scales, from the bandwidth $D$ down to the Kondo temperature $T_K$.

In the NH case, $\hat{H}_N$ in general has complex eigenvalues, and its left and right eigenvectors are distinct. The iterative diagonalization procedure in NH-NRG proceeds similarly to the Hermitian case, but the recursion by which $\hat{H}_{N+1}$ is obtained from $\hat{H}_N$ must be carefully reformulated to account for these crucial differences -- see \textit{End Matter} and \cite{SuppMat}. 
One may construct a bi-orthonormal basis~\cite{Brody_BiorthogonalQM_JoPA2014} if the spectrum is non-degenerate, and this provides substantial advantages in terms of the efficiency and stability of the algorithm. Although quantum impurity models do typically have many eigenvalue degeneracies, the most significant source of these is from \textit{symmetries}. However, these symmetries can then be utilized to block-diagonalize the Hamiltonian in distinct quantum number subspaces~\footnote{Non-abelian symmetries can also be leveraged to diagonalize the Hamiltonian in \textit{multiplet} space~\cite{weichselbaum2007sum}}. In the present setting, labeling states by the total charge $Q$ and total spin projection $S_z$ is sufficient to separate all exact degeneracies into different blocks~\cite{lee2016adaptive}. It is anyway desirable to exploit symmetries in this way since it reduces block sizes, and increases computational efficiency~\cite{weichselbaum2012non}. We identify two other sources of approximate degeneracy in these systems: accidental and emergent. In both cases, the use of high-precision numerics is found to overcome any instabilities associated with bi-orthonormalization~\cite{SuppMat}.

Another key aspect of the NRG procedure that must be adapted is the Fock-space truncation at each step. In Hermitian NRG, where the eigenvalues $E_N$ are real, we retain only the $N_k$ lowest-lying eigenvalues; but this becomes ambiguous in the NH context when the eigenvalues are complex. We found that truncating by the lowest real-part of the eigenvalues gives the most accurate and stable results. We therefore identify the `ground state' as the one with the lowest real-part (consistent with existing conventions in NH physics). 

We have confirmed explicitly that applying NH-NRG to a non-interacting NH resonant level model using this truncation scheme perfectly reproduces the results of exact diagonalization, as shown in the \textit{End Matter}. This provides a stringent test of the NH-NRG algorithm.  

Our NH-NRG code is available open source to facilitate future studies of NH quantum impurity models, see~\cite{GitHub}.

%%%%%%%%%%%%%%%%%

\textit{Solution of the NH Kondo model.--} We now apply NH-NRG to the metallic NH Kondo model (bandwidth $D\equiv 1$ hereafter). The phase diagram for antiferromagnetic $J_R>0$ obtained by NH-NRG is presented in Fig.~\ref{fig_Kondo_PhaseDiagram}(a) as a function of the real and imaginary parts of the complex Kondo coupling, $J_R$ and $J_I$. We find two phases, described by the SC and LM fixed points of the \textit{Hermitian} Kondo model, separated by a %first-order 
quantum phase transition. We identify the phases from the NH-NRG eigenspectrum at large $N$ after convergence, which takes a distinct form in SC and LM phases. In particular, the imaginary part of the eigenvalues ${\rm Im}(E_N)$ vanish in all cases at large $N$, indicating the emergent Hermiticity of the fixed point Hamiltonian. Since the fixed points are Hermitian, we can compute their thermodynamic properties in the usual way~\cite{wilson1975renormalization}. As expected, we find an impurity contribution to entropy of $k_B\ln(2)$ for a free spin in the LM phase, and $0$ for the screened Kondo singlet in the SC phase.

At relatively weak bare coupling $|J| \lesssim 0.25$, the NH-NRG phase boundary (black line) matches precisely with the Bethe ansatz prediction of Ref.~\cite{NakagawaKawakamiUeda_NonHermKondo_PRL2018}, plotted as the red dot-dashed line (see inset). However at stronger coupling we find new features. For $J_R \gtrsim 0.55$ the LM phase disappears, and the Kondo effect dominates over dissipative effects. For $J_R \lesssim 0.55$ we find re-entrant Kondo physics as $|J_I|$ is increased. Therefore, the dissipation-induced unscreened phase in fact occupies a bounded region in the parameter space of the NH Kondo model. Interestingly, similar phase diagrams have been observed in other non-Hermitian many-body systems~\cite{yamamoto2019theory,nakagawa2021exact,li2023yang}.

We analyze the RG flow in Fig.~\ref{fig_Kondo_PhaseDiagram}(b-e) by tracking the (rescaled) NRG eigenvalues $E_N$ as a function of iteration number $N$. In (b) we plot the real and imaginary parts (top and bottom panels) for a system in the SC phase, and observe clear RG flow between LM and SC fixed points. Although the imaginary part of $E_N$ is finite for early iterations and initially grows, it decays to zero as the stable fixed point is reached. Interestingly, ${\rm Im}(E_N)$ becomes large along the \textit{crossover} between fixed points. The crossover at $N_c$ between LM and SC can be interpreted as a `temperature' scale $T_K\sim D\Lambda^{-N_c/2}$ corresponding to Kondo screening, and we numerically extract the relation $T_K\sim D e^{-2DJ_R/|J|^2}$ from the NH-NRG data in the weak-coupling regime~\cite{SuppMat}, consistent with the perturbative scaling result of Ref.~\cite{NakagawaKawakamiUeda_NonHermKondo_PRL2018}.

Fig.~\ref{fig_Kondo_PhaseDiagram}(c) shows the analogous plots for a system in the LM phase, which starts off close to the LM fixed point, evolves under RG, but then returns back to it at large $N$. This anomalous RG-flow reversion, identified in Ref.~\cite{NakagawaKawakamiUeda_NonHermKondo_PRL2018}, is further illustrated in panels (d,e) which show the evolution of two particular eigenvalues in the Argand plane, with increasing $N$ following the direction of the arrows. Green points show the fixed point eigenvalues of the Hermitian Kondo model to which they converge.

The phase transition is controlled by a \textit{non-Hermitian} critical fixed point~\cite{SuppMat}. At the critical point $J=J_c$,  ${\rm Im}(E_N)$ diverges exponentially with $N$. Near the critical point, we identify a scale that vanishes as $T^*\sim |J-J_c|$.

%%%%%%%%%%%%%%%%%

\textit{NH Pseudogap Kondo.--} To further showcase the versatility of the NH-NRG method, we now turn to the NH pseudogap Kondo model. The pseudogap bath is characterized by a density of states $\rho(\omega)=\rho_0|\omega|^r\Theta(D-|\omega|)$ with power-law exponent $r>0$, and we focus on the particle-hole symmetric case. The standard Hermitian version of the model has been extensively studied using a variety of methods including perturbative RG~\cite{fritz2004phase,FritzVojta_PseudogapCritical_PRB2006} and NRG~\cite{Bulla_AIM_Pseudogap_JoP1997,gonzalez1998renormalization}. A transition between LM and SC phases upon increasing $J_R$ through the critical value $J_R^*(r)$ was found for $0<r<\tfrac{1}{2}$, with $r=\tfrac{1}{2}$ itself playing the role of a lower-critical dimension $r_c$, beyond which the critical point disappears and  Kondo screening is no longer possible~\cite{fritz2004phase}. By contrast, for the NH variant with $J \in \mathbb{C}$ we find that $r_c\equiv r_c(J_I)$ gets shifted to larger values as $J_I$ increases. Fig.~\ref{fig_PseudoKondo_PhaseDiagram}(a) shows the phase transition boundaries obtained from NH-NRG as a function of $J_R$ and $r$ for different $J_I$. The blue line is for the Hermitian case with $J_I=0$, which is seen to diverge at $r_c(0)=\tfrac{1}{2}$ as expected from Ref.~\cite{Bulla_AIM_Pseudogap_JoP1997}. For $J_I>0$ and  $0<r<\tfrac{1}{2}$ our analysis of the eigenvalue RG flow shows that the stable fixed points obtained at large $N$ are identical to the Hermitian pseudogap Kondo fixed points. Fig.~\ref{fig_PseudoKondo_PhaseDiagram} shows the flow diagrams for $J_R<J_R^*$ in the LM phase (panel b) and for $J_R>J_R^*$ in the SC phase (c). Likewise, the LM phase for $r>\tfrac{1}{2}$ in panel (d) shows RG flow to the standard Hermitian LM fixed point. However, in the region $r>\tfrac{1}{2}$ and $J_R>J_R^*$ that would be forbidden in the Hermitian limit, we find an entirely novel stable fixed point, see panel (e). Remarkably, in this phase the stable fixed point is intrinsically non-Hermitian, with a persistent complex eigenspectrum and ${\rm Im}(E_N)$ that do not decay with $N$. We dub this fixed point CSC for complex strong coupling. We leave the detailed study of this phase to future work.
This behavior and the structure of the full phase diagram is beyond the reach of perturbative techniques~\cite{chen2024criticalbehaviornonhermitiankondo} or methods relying on linear dispersion.

%%%%%%%%%%%%%%

%--\FIGURE\---------------
\begin{figure}[t!]
	\centering 
    \includegraphics[width=\linewidth]{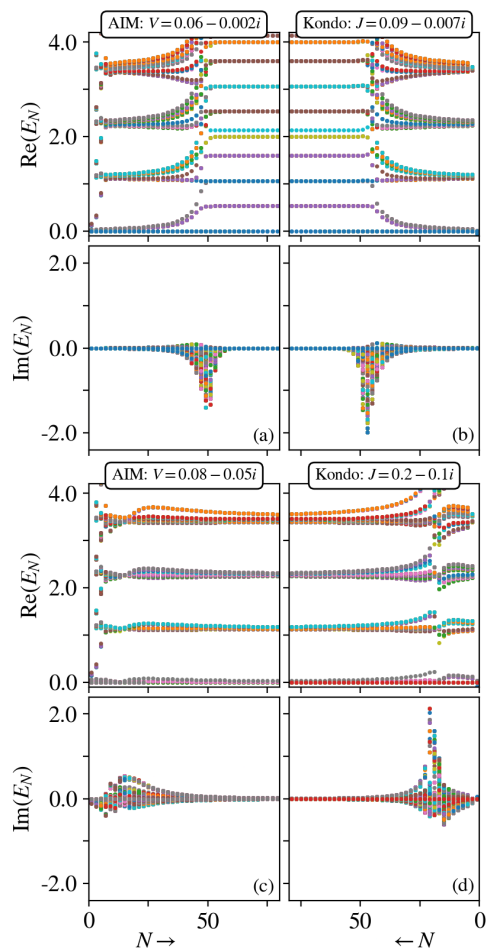}
    \caption{Comparison of eigenvalue RG flows for NH AIM (left) and NH Kondo (right). (a,b)  Convergence to the same SC fixed point; (c,d) convergence to the LM fixed point. NH-NRG calculations for $U_d = 0.3$, $\epsilon_d=-0.15$, $\Lambda=3$, $N_k=400$. 
    }
	\label{fig_AIM_flow}
\end{figure}
%--\FIGURE\-------------

\textit{Non-Hermitian Anderson model.--} Finally, we consider the physics of the NH Anderson impurity model (AIM),
\begin{eqnarray}
    \hat{H}_{\rm AIM} =& \hat{H}_{\rm bath} + \epsilon_d \sum_{\sigma}d^{\dagger}_{\sigma}d_{\sigma}^{\phantom{\dagger}} + U_d d^{\dagger}_{\uparrow}d_{\uparrow}^{\phantom{\dagger}}d^{\dagger}_{\downarrow}d_{\downarrow}^{\phantom{\dagger}} \nonumber \\    &+ V\sum_{\sigma} \left( d^{\dagger}_{\sigma}c_{0\sigma}^{\phantom{\dagger}} + c^{\dagger}_{0\sigma}d_{\sigma}^{\phantom{\dagger}} \right) \:,
    \label{eq_AIM_Ham}
\end{eqnarray}
where the first line describes the isolated bath and impurity orbital, while the tunnel-coupling between them is given in the second line. Non-Hermiticity can be introduced by making any/all of the parameters $\epsilon_d$, $U_d$ or $V$ complex.  We focus here on the case where $V\in \mathbb{C}$ and the bath has a flat density of states. Various aspects of Anderson models describing loss and dephasing have been considered before~\cite{Lourenco_PTKondo_PRB2018, Kulkarni_PTAndersonModel_PRB2022, yamamoto2024correlationversusdissipationnonhermitian,kulkarnianderson,vanhoecke2024kondozenocrossoverdynamicsmonitored}, but our aim here is to confirm the mapping between NH AIM and Kondo models. The Schrieffer-Wolff transformation~\cite{Hewson1993,Lourenco_PTKondo_PRB2018,vanhoecke2024kondozenocrossoverdynamicsmonitored} is perturbative and applies strictly only in the limit of large $U_d$ (and therefore small $J$). Is the `low-energy' physics, and especially the ground state, of the AIM described by the Kondo model beyond the perturbative regime? Is the phase diagram of the NH Kondo model shown in Fig.~\ref{fig_Kondo_PhaseDiagram}(a) accessible within the AIM?

We answer these questions using non-perturbative NH-NRG. The mapping between Hermitian AIM and Kondo models beyond Schrieffer-Wolff was first established in Ref.~\cite{KrishnamurthyWilson_NRG1_PRB1980} using NRG, and we adopt the same strategy here for the NH case. In Fig.~\ref{fig_AIM_flow} we confirm  explicitly that the same stable fixed points are reached in the same way under RG in both models, for both SC and LM phases~\footnote{Note that the initial flow for the AIM is expected to be different from that of the Kondo model because of the existence of the additional `free orbital' fixed point in the UV.}.  
NH-NRG results show that the phase diagram of the NH AIM in the $({\rm Re}V,{\rm Im}V)$ plane has the same structure as that of the NH Kondo model, including the re-entrant Kondo behavior at large ${\rm Im}V$ \cite{SuppMat} and termination of the LM phase beyond a critical value of ${\rm Re}V$. Ferromagnetic $J_R<0$ is not accessible in the NH Kondo model.

%%%%%%%%%%%%%%

\textit{Conclusion and outlook.--}
The numerical renormalization group is often considered the gold-standard method of choice for solving quantum impurity models~\cite{Bulla_NRG_RevMod2008}. Here we generalized the method to treat non-Hermitian impurity problems, and applied our NH-NRG approach to the NH Kondo and NH Anderson models. NH-NRG is non-perturbative and can be applied equally well to non-integrable systems and those without the linear dispersion property, such as the pseudogap Kondo model. The method provides direct access to the RG flow of the complex many-particle eigenvalues: it allows different phases to be fingerprinted by identification of characteristic fixed point structures, and emergent energy scales can be read off from the crossovers between fixed points.

%Our numerically-exact results clarify the phase diagram of the NH Kondo model beyond weak coupling, showing that the dissipation-induced local-moment phase occupies a bounded region in parameter space, with re-entrant Kondo screening arising for strong dissipation. We show that the low-energy properties of the NH Anderson model are described by the NH Kondo model, and the full phase diagram of the latter is accessible in the former. In the pseudogap NH Kondo model, a novel NH fixed point is identified with complex eigenvalues.

NH-NRG opens the door to studying the interplay between NH and strong correlation physics in a wide range of models -- for example systems with multiple impurities~\cite{affleck1995conformal,allerdt2015kondo,mitchell2009quantum,vzitko2006multiple,karki2023z} and/or multiple baths~\cite{nozieres1980kondo,toth2007dynamical,iftikhar2018tunable,mitchell2012two}, impurities in unconventional materials~\cite{fritz2013physics,mitchell2013kondo,shankar2023kondo,mitchell2013kondo2,mitchell2015kondo,galpin2008anderson,moca2021kondo,vojta2016kondo}, underscreened Kondo effect with higher spin~\cite{parks2010mechanical,mehta2005regular}, and critical phenomena near impurity quantum phase transitions~\cite{affleck1991critical,vojta2006impurity}. NH-NRG could be extended to compute zero-temperature dynamical quantities, such as the impurity spectral function~\cite{weichselbaum2007sum}. This would allow non-Hermitian lattice models~\cite{Fukui_AsymmHubb_PRB1998,Hayata_AsymmHubbard_PRB2021} to be studied within DMFT~\cite{georges1996dynamical}, using NH-NRG as an impurity solver. Our NH-NRG code is provided open source at Ref.~\cite{GitHub}.\\

%%%%%%%%%%%%%%%%%%%%%%%%%%%%%%%%%%%%%%%%%%%%%%%%%%%%%%%%%%%%%%%%%%%%%%%%%%%%%%%%%%%%%%%%%%
\begin{acknowledgments}
\textit{Acknowledgments.--} The authors acknowledge funding from Science Foundation Ireland through grant 21/RP-2TF/10019. We are grateful to Ralph Smith for providing the `GenericSchur’ Julia package~\cite{shur}.
\vfill
\clearpage

\appendix
%%%%%%%%%%%%%%%%%%%%%%%%%%%%%%%%%%%%%%%%%%%%%%%%%%%%%%%%%%%%%%%%%%%%%%%%%%%%%%%%%%%%%%%%%%
\section*{End Matter}

%%%%%%%%%%%%%%%%%%%%%%%%%%%%%%%%%%%%%%%%%%%%%%%%%%%%%%%
% Counter resetting for supplementary

%\setcounter{page}{1} \renewcommand{\thepage}{E\arabic{page}}

\setcounter{figure}{0}
\renewcommand{\thefigure}{E\arabic{figure}}

\setcounter{equation}{0}
\renewcommand{\theequation}{E.\arabic{equation}}

\setcounter{table}{0}
\renewcommand{\thetable}{E.\arabic{table}}

\setcounter{section}{0}
\renewcommand{\thesection}{E.\Roman{section}}

\renewcommand{\thesubsection}{E.\Roman{section}.\Alph{subsection}}

\makeatletter
\renewcommand*{\p@subsection}{}
\makeatother

\renewcommand{\thesubsubsection}{E.\Roman{section}.\Alph{subsection}-\arabic{subsubsection}}

\makeatletter
\renewcommand*{\p@subsubsection}{}  % referring to subsubsections 
\makeatother

%%%%%%%%%%%%%%%%%%%%%%%%%%%%%%%%%%%%%%%%%%%%%%%%%%%%%%%%%%%%%%%%%%%%%%%%%%%%%%%%%%%%%%%%%%

\textit{Iterative diagonalization.--} 
Here we give an overview of the NH-NRG algorithm, highlighting  key differences with the Hermitian formulation described in Ref.~\cite{Bulla_NRG_RevMod2008}. In the following we assume a bi-orthonormal basis~\cite{Brody_BiorthogonalQM_JoPA2014} such that the inner product of left and right states satisfies $\innerprod{n}{m}^{L,R}=\delta_{nm}$. Further details can be found in the Supplemental Material~\cite{SuppMat}. 

At step $N+1$ of the NH-NRG calculation, we construct the Hamiltonian matrix $\mathbf{H}_{N+1}^b$ with elements,
\begin{eqnarray}
	\bra{N+1;k;r}_{b}^{L} \hat{H}_{N+1}^{\phantom{\dagger}} \ket{N+1;k';r'}_{b}^{R} \;,
    \label{eq_HamTerm} 
\end{eqnarray}
where $L$($R$) refers to the left(right) states, and the $b$ subscript denotes the \textit{basis} states, which are decomposed as,
\begin{eqnarray}\label{eq:decomp}
	\ket{N+1;k;r}_{b}^{L(R)} = \ket{k}_{N+1} \otimes \ket{N;r}_{d}^{L(R)} \;.
\end{eqnarray}
Here $\ket{N;r}_{d}^{L(R)}$ are the $N_k$ retained left(right) \textit{eigenstates} of the previous iteration satisfying $\hat{H}_N\ket{N;r}_{d}^{R}=E_N(r)\ket{N;r}_{d}^{R}$ and $\bra{N;r}_d^L\hat{H}_N=\bra{N;r}_{d}^{L}E_N(r)$; whereas $\ket{k}_{N+1}=\{\ket{-},\ket{\downarrow},\ket{\uparrow},\ket{\uparrow\downarrow} \}$ are the four states of the added orbital $N+1$, labeled respectively by the index $k=\{0,-1,+1,2\}$, which are equal for $L$ and $R$.

From the recursion relation $\hat{H}_{N+1}=\hat{H}_N+\hat{T}_{N+1}$ and Eq.~\eqref{eq:decomp}, we may then express the matrix elements as,
\begin{eqnarray}
&&\bra{N+1;k;r}_{b}^{L}\hat{H}_{N+1}^{\phantom{\dagger}}\ket{N+1;k';r'}_{b}^{R}  \\
&&= \bra{k}_{N+1}\bra{N;r}_{d}^{L} \hat{H}_N \ket{k'}_{N+1}\ket{N;r'}_{d}^{R} \nonumber \\
&&+ \epsilon_{N+1} \sum_{\sigma} \bra{k}_{N+1}\bra{N;r}_{d}^{L} f_{N+1\sigma}^{\dagger} f_{N+1\sigma}^{\phantom{\dagger}}  \ket{k'}_{N+1}\ket{N;r'}_{d}^{R} \nonumber \\
&&+ t_N \sum_{\sigma} \bra{k}_{N+1}\bra{N;r}_{d}^{L} f_{N\sigma}^{\dagger} f_{N+1\sigma}^{\phantom{\dagger}} \ket{k'}_{N+1}\ket{N;r'}_{d}^{R} \nonumber \\
&&+ t_N \sum_{\sigma} \bra{k}_{N+1}\bra{N;r}_{d}^{L} f_{N+1\sigma}^{\dagger} f_{N\sigma}^{\phantom{\dagger}}  \ket{k'}_{N+1}\ket{N;r'}_{d}^{R} \;. \nonumber
\end{eqnarray}
This expression simplifies to,
\begin{eqnarray}
&&\bra{N+1;k;r}_{b}^{L}\hat{H}_{N+1}^{\phantom{\dagger}}\ket{N+1;k';r'}_{b}^{R}  \\
&&= \delta_{kk'}\delta_{rr'}\left(E_N(r)+|k|\epsilon_{N+1}  \right)  \nonumber \\
&&+ (-1)^k \:t_N \sum\nolimits_{\sigma} M^\sigma_{kk'} \bra{N;r}_{d}^{L} f_{N\sigma}^\dagger \ket{N;r'}^R_d \nonumber \\
&&+ (-1)^{k'} \:t_N \sum\nolimits_{\sigma} M^\sigma_{k'k} \bra{N;r}_{d}^{L} f_{N\sigma} \ket{N;r'}^R_d \nonumber 
\end{eqnarray}
where in the last two lines we inserted the identity between the creation and annihilation operators~\cite{SuppMat}. Here, $M^{\sigma\phantom{\dagger}}_{kk'}=(M^{\sigma}_{k'k})^{\dagger}$ denotes the trivial matrix element $\langle{k}|f_{N+1\sigma}|{k'}\rangle_{N+1}$ whose value does not depend on $N$.

Thus we can construct the Hamiltonian matrix $\mathbf{H}_{N+1}^b$ at NRG iteration $N+1$ using information from iteration $N$. Specifically, we need the set of complex eigenvalues $E_N(r)$, and the matrix elements $\bra{N;r}_{d}^{L} f_{N\sigma}^\dagger \ket{N;r'}^R_d$ and $\bra{N;r}_{d}^{L} f_{N\sigma} \ket{N;r'}^R_d$ which in the NH case are \textit{not} Hermitian conjugates and need to be computed separately.

With $\mathbf{H}_{N+1}^b$ in hand, we diagonalize the matrix to obtain the eigenvalues $E_{N+1}$ and the left and right eigenvectors $\ket{N+1;r}_{d}^{L(R)}$. Specifically, $\mathbf{H}_{N+1}^b=\mathbf{U}_{N+1}^R \mathbf{H}_{N+1}^d (\mathbf{U}_{N+1}^L)^{\dagger}$ where  $\mathbf{H}_{N+1}^d$ is the \textit{diagonal} matrix of eigenvalues $E_{N+1}$ and  $\mathbf{U}_{N+1}^{R(L)}$ is a matrix whose columns are the right(left) eigenvectors. Therefore we can expand the eigenstates as,
\begin{gather}
	\ket{N+1;r}_{d}^{R(L)} = \sum_{m,s}U_{N+1}^{R(L)}(r;m,s)^{(\dagger)} \ket{N+1;m;s}_{b}^{R(L)} \nonumber\\
    \equiv \sum_{m,s}U_{N+1}^{R(L)}(r;m,s)^{(\dagger)} \ket{m}_{N+1}\ket{N;s}_{d}^{R(L)} \;.
\end{gather}
We use this to construct the nontrivial matrix elements required for the next step,
\begin{eqnarray}
&&\bra{N+1;r}_{d}^{L} f_{N+1\sigma}^\dagger \ket{N+1;r'}^R_d  \\ 
   &&= \sum_{m,m',s}  M^\sigma_{m'm} \:U^{L}_{N+1}(r;m,s)^{\dagger}  U^{R}_{N+1}(r';m',s)  \nonumber \\
   &&\bra{N+1;r}_{d}^{L} f_{N+1\sigma} \ket{N+1;r'}^R_d \\ 
   &&= \sum_{m,m',s}  M^\sigma_{mm'} \:U^{L}_{N+1}(r;m,s)^{\dagger}  U^{R}_{N+1}(r';m',s)  \nonumber
\end{eqnarray}
Note that only the `lowest' $N_k$ eigenstates are retained at each step, meaning that the computational complexity is approximately constant at each step. In practice this Fock space truncation is done by retaining states with the lowest \textit{real part} of the complex eigenvalues $E_N$. 

As such, the chain can be built up iteratively, starting from $\hat{H}_0$ consisting of just the impurity and the Wilson zero-orbital. Since states with large ${\rm Re}(E_N)$ are discarded at each step, we focus on the states with progressively smaller ${\rm Re}(E_N)$ as the calculation proceeds. To analyze the RG flow we specify $E_N$ with respect to the state with the lowest ${\rm Re}(E_N)$ at that iteration, and rescale by a factor of $\Lambda^{N/2}$. It is these rescaled eigenvalues that are plotted in the figures.

%%%%%%%%%%%%%%%%%%%

%--\FIGURE\---------------------
\begin{figure*}[t!]
	\centering 
    \includegraphics[width=\linewidth]{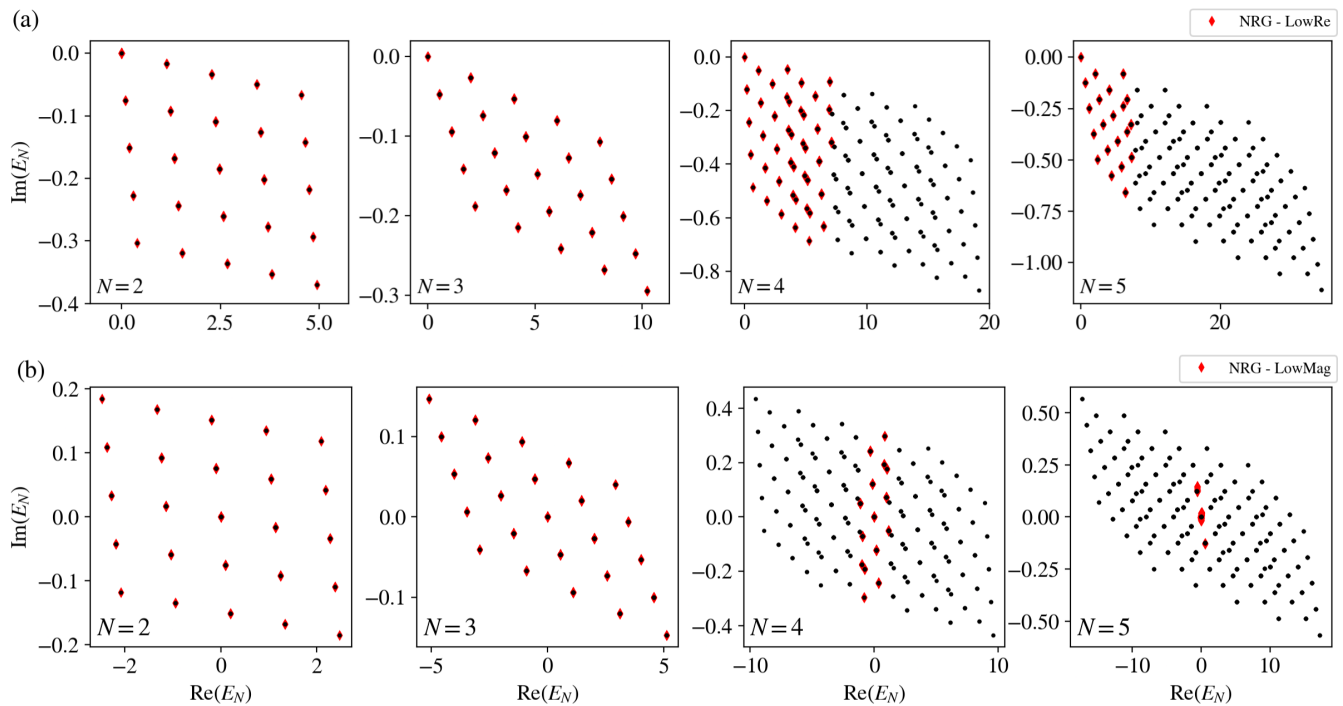}
 \caption{Validation of NH-NRG method for the non-interacting AIM with $U_d=0$.  The full set of complex eigenvalues are constructed from exact diagonalization of $\hat{H}_N$ and shown in the Argand plane for $N=2,3,4,5$ as the black circle points. NH-NRG results are shown as the red diamond points: the ${\rm min}(N_k,4^{N+2})$ retained states match precisely with the exact results. Top row (a) shows NH-NRG truncation scheme `LowRe' in which the lowest $N_k$ states sorted by ${\rm Re}(E_N)$ are kept. Bottom row (b) shows an alternative truncation scheme `LowMag' where states are sorted by $|E_N|$. Shown for $\Lambda=3$, $N_k=1024$, $\epsilon_d=0$, $V=0.1-0.08i$. Eigenvalues rescaled by $\Lambda^{N/2}$ are plotted with respect to the `ground state' of that iteration.
    }
	\label{fig_tb_lowre}
\end{figure*}
%--\FIGURE\----------------------

\textit{Truncation schemes and numerical precision.--}
Through extensive numerical testing, we found that truncation to the $N_k$ states with the lowest ${\rm Re}(E_N)$ at each step yields the most stable and accurate results. In this Letter we presented results for $\Lambda=3$ and $N_k=400$, which we explicitly checked were numerically converged with respect to increasing $N_k$ (essentially no change in the RG flow was observed by increasing $N_k$ to $1024$ kept states). In certain cases we observed numerical instabilities in the diagonalization which were completely resolved by using high-precision numerics. All of the results presented were confirmed to be converged using 128-bit precision complex numbers~\cite{shur}.

Other truncation schemes (discussed further below) were investigated. For example, truncation to the $N_k$ states with the lowest \textit{magnitude} $|E_N|$ produces a somewhat different set of states being tracked along the RG flow. However, retained states common to both truncation schemes were found to have exactly the same RG flow, provided $N_k$ was sufficiently large. Overall, truncation by lowest ${\rm Re}(E_N)$ is preferred due to advantageous stability and accuracy with respect to $N_k$, and less frequent need for high-precision numerics.

%%%%%%%%%%%%%%%%%

\textit{Validation and benchmarking of method.--}
The NH-NRG method for the AIM works with equal ease for any interaction strength $U_d$. In particular, the NH-NRG algorithm works exactly the same for the trivial case $U_d=0$ as for the nontrivial interacting case with $U_d>0$. For $U_d=0$ we can also simply diagonalize the single-particle Hamiltonian matrix and then construct the many-particle states from these. Thus in this limit we can exactly diagonalize the full impurity-and-Wilson-chain composite system without any truncation or approximation. This provides a stringent check on our NH-NRG results by direct comparison.

Our results of this testing are shown in Fig.~\ref{fig_tb_lowre} for the $U_d=0$ AIM (also known as the resonant level model) in which a non-interacting impurity is coupled to the usual (flat-band) Wilson chain of $N+1$ sites. We compare NH-NRG results (red diamond points) with $N_k=1024$ kept states, and exact diagonalization of the tight-binding model (black circle points). In the latter we construct the full $4^{N+2}$ dimensional Fock space. Complex eigenvalues of $\hat{H}_N$ are plotted for different $N$ in the Argand plane in Fig.~\ref{fig_tb_lowre}. Top row (a) shows results for the truncation scheme `LowRe' in which the $N_k$ states with the lowest ${\rm Re}(E_N)$ are retained; whereas the bottom row (b) is for the `LowMag' scheme where the $N_k$ states with the lowest $|E_N|$ are instead kept. Although a different set of states in the NH-NRG calculation is retained in either case, these accurately match with the corresponding exact diagonalization results from the tight-binding chain. We note that the NH-NRG results for $N=5$ in panel (b) are highly degenerate, with the $N_k=1024$ retained states giving only three distinct eigenvalues. The kept eigenvalues in (a) are far less degenerate. 
In the case of high degeneracy, which poses a challenge for numerical diagonalization and bi-orthonormalization of NH matrices, we add a physically inconsequential on-site disorder to the Wilson chain of width $10^{-7}$ which lifts the degeneracy. This precaution was not required for any of the interacting models studied, which do not possess such high degeneracies.

Additional examples of different truncation schemes and benchmarking for different non-interacting models are provided in the Supplemental Material~\cite{SuppMat}. We have also checked that our NH-NRG code reproduces the results of standard NRG when $\hat{H}_0$ is Hermitian. 

\clearpage

\end{acknowledgments}

%%%%%%%%%%%%%%%%%%%%%%%%%%%%%%%%%%%%%%%%%%%%%%%%%%%%%%%%%%%%%%%%%%%%%%%%%%%%%%%%%%%%%%%%%%
%\bibliographystyle{unsrt} % unsrt, acmapsrev4-1, apsrev4-1
%\bibliography{Kondo_bib}

%apsrev4-2.bst 2019-01-14 (MD) hand-edited version of apsrev4-1.bst
%Control: key (0)
%Control: author (8) initials jnrlst
%Control: editor formatted (1) identically to author
%Control: production of article title (0) allowed
%Control: page (0) single
%Control: year (1) truncated
%Control: production of eprint (0) enabled
%

\clearpage
%%%%%%%%%%%%%%%%%%%%%%%%%%%%%%%%%%%%%%%%%%%%%%%%%%%%%%%%%%%%%%%%%%%%%%%%%%%%%%%%%%%%%%%%%%

%%%%%%%%%%%%%%%%%%%%%%%%%%%%%%%%%%%%%%%%%%%%%%%%%%%%%%%
% Counter resetting for supplementary

\setcounter{page}{1} \renewcommand{\thepage}{S\arabic{page}}

\setcounter{figure}{0}
\renewcommand{\thefigure}{S\arabic{figure}}

\setcounter{equation}{0}
\renewcommand{\theequation}{S.\arabic{equation}}

\setcounter{table}{0}
\renewcommand{\thetable}{S.\arabic{table}}

\setcounter{section}{0}
\renewcommand{\thesection}{S.\Roman{section}}

\renewcommand{\thesubsection}{S.\Roman{section}.\Alph{subsection}}

\makeatletter
\renewcommand*{\p@subsection}{}
\makeatother

\renewcommand{\thesubsubsection}{S.\Roman{section}.\Alph{subsection}-\arabic{subsubsection}}

\makeatletter
\renewcommand*{\p@subsubsection}{}  % referring to subsubsections 
\makeatother

%%%%%%%%%%%%%%%%%%%%%%%%%%%%%%%%%%%%%%%%%%%%%%%%%%%%%%%%%%%%%%%%%%
%% Abstract etc. %%%%%%%%%%%%%%%%%%%%%%%%%%%%%%%%%%%%%%%%%%%%%%%%%
%%%%%%%%%%%%%%%%%%%%%%%%%%%%%%%%%%%%%%%%%%%%%%%%%%%%%%%%%%%%%%%%%%
\onecolumngrid
	\begin{center}
		\large{
			Supplemental Material for \\ \textit{Non-Hermitian Numerical Renormalization Group:\\ Solution of the non-Hermitian Kondo model}
		}
	\end{center}  
	
	%\author{Phillip C. Burke}
	%\email[]{phillip.cussenburke@ucd.ie}
	%\affiliation{School of Physics, University College Dublin, Belfield, Dublin 4, Ireland}
	%\affiliation{Centre for Quantum Engineering, Science, and Technology, University College Dublin, Dublin 4, Ireland}
	
	%\author{Andrew K. Mitchell} 
	%\affiliation{School of Physics, University College Dublin, Belfield, Dublin 4, Ireland}
	%\affiliation{Centre for Quantum Engineering, Science, and Technology, University College Dublin, Dublin 4, Ireland}

	\maketitle
	%%%%%%%%%%%%%%%%%%%%%%%%%%%%%%%%%%%%%%%%%%%%%%%%%%%%%%%%%%%%%%%%%%%%%%%%%%%%%%%.
	
	%\begin{widetext}
	
	%%%%%%%%%%%%%%%%%%%%%%%%%%%%%%%%%%%%%%%%%%%%%%%%%%%%%%%%%%%%%%%%
	%% Overview %%%%%%%%%%%%%%%%%%%%%%%%%%%%%%%%%%%%%%%%%%%%%%%%%%%%
	
	In this Supplemental Material we provide supporting information and data. 
	\begin{itemize}
		\item In Section \ref{sec:NH}, we discuss some basic properties of non-Hermitian (NH) systems.
		\item In Section \ref{app_IterativeConstruction}, we provide the complete derivation of the iterative construction of the Hamiltonian used in the non-Hermitian numerical renormalization group (NH-NRG) method.
		\item In Section \ref{app_alt_truncs}, we illustrate alternative truncation schemes for the NH-NRG procedure.
		\item In Section \ref{app_AIM}, we provide additional eigenvalue flow diagrams for the NH Anderson Impurity Model (AIM).
		\item In Section \ref{app_Tk}, we discuss the evolution of the Kondo temperature $T_K$.
		\item In Section \ref{app_crit}, we present further results for the critical point of the NH Kondo model.
	\end{itemize} 
	
	%%%%%%%%%%%%%%%%%%%%%%%%%%%%%%%%%%%%%%%%%%%%%%%%%%%%%%%%%%%%%%%%
	
	\section{Non-Hermitian systems}\label{sec:NH}
	
	Before jumping into the iterative construction procedure used in NH-NRG, we first provide a brief discussion of NH matrices which will come in useful later. See also Refs.~\cite{supp_Brody_BiorthogonalQM_JoPA2014, supp_Edvardsson_Biorthogonal_arxiv2023} for discussions of bi-orthogonal quantum mechanics.
	
	For an NH system, for which $\hat{H} \neq \hat{H}^\dagger$, the left and right eigenvectors are defined such that,
	\begin{eqnarray}
		\hat{H}\ket{E_j}^{R} = \lambda_j\ket{E_j}^{R} \ ~~ &,&~~ \ \hat{H}^\dagger\ket{E_j}^{L} = \lambda^*_j\ket{E_j}^{L} \\
		\bra{E_j}^{R}\hat{H}^\dagger = \bra{E_j}^{R}\lambda^*_j \ ~~ &,& ~~ \ \bra{E_j}^{L}\hat{H} = \bra{E_j}^{L}\lambda_j.
	\end{eqnarray}  
	Although the left and right eigenvectors are not individually orthonormal, they may form a bi-orthogonal basis if the eigenspectrum is non-degenerate. In the following we assume this property, which can be defined as,
	\begin{eqnarray}
		\innerprod{E_i}{E_j}^{LR} =  \delta_{ij} \;.
	\end{eqnarray}
	Here we have also bi-\textit{normalized} the basis. 
	We note that a bi-orthonormal basis is not the default output for left and right eigenvectors from most standard numerical eigensolvers (e.g.~via Python, Julia, or Fortran) and so the bi-normalization typically has to be done manually.
	
	To bi-normalize the left and right eigenvectors, we first compute the overlaps,
	\begin{gather}
		\operatorname{LR}_j = \innerprod{E_j}{E_j}^{LR} \;,
	\end{gather}
	and then, provided the corresponding left and right eigenvectors are non-orthogonal, we rescale the vectors, 
	\begin{gather}
		|E_j\rangle^{R} \to \frac{\ket{E_j}^{R}}{\sqrt{\operatorname{LR_j}}} \qquad , \qquad 
		|E_j\rangle^{L} \to \frac{\ket{E_j}^{L}}{\sqrt{\operatorname{LR_j}}^*} \;,
		\label{eq_binorm}
	\end{gather}
	which ensures $\innerprod{E_j}{E_j}^{LR} =  1$.
	
	Assuming bi-orthonormality now, an NH matrix $\hat{H}$ can be decomposed in terms of its left and right eigenvectors,
	\begin{gather}
		\hat{H}^{\phantom{N}} = \sum_j \lambda_j \outerprod{E_j}{E_j}^{RL}  \qquad , \qquad 
		\bra{E_j}^{L} \hat{H} \ket{E_k}^{R} = \delta_{jk} \lambda_j \;.
	\end{gather}
	
	%%%%%%%%#
	
	With some bi-normalized basis $\ket{\phi_j}^{L(R)}$ of left(right) states, we can construct the Hamiltonian matrix $\mathbf{H}_{\phi}$ with elements $[\mathbf{H}_{\phi}]_{ij}=\bra{\phi_i}^L\hat{H}\ket{\phi_j}^R$. Numerical diagonalization of this matrix yields $\mathbf{U}^R\: \mathbf{H}_{E}\: (\mathbf{U}^L)^{\dagger}=\mathbf{H}_{\phi}$ where $[\mathbf{H}_{E}]_{ij}=\delta_{ij}\lambda_j$ and the columns of the matrices $\mathbf{U}^R$ and $\mathbf{U}^L$ contain the right and left eigenvectors. It follows that,
	\begin{gather}
		(\mathbf{U}^L)^{\dagger} \mathbf{U}^R=\mathbf{I} \qquad ; \qquad \tr[(\mathbf{U}^L)^{\dagger} \mathbf{U}^R]= \dim(\mathbf{H}) \;.
	\end{gather} 
	However, note that $(\mathbf{U}^{L(R)})^{\dagger} \mathbf{U}^{L(R)} \ne \mathbf{I}$ since the left and right sets themselves are not orthonormal.
	
	Importantly, for bi-orthonormal systems the identity can be resolved as,
	\begin{equation}
		\mathbb{1}=\sum_j \ket{\phi_j}\bra{\phi_j}^{RL} \;.
	\end{equation}
	
	One issue with the bi-orthonormalization procedure is that it requires a non-degenerate eigenspectrum~\cite{supp_Brody_BiorthogonalQM_JoPA2014}.  In general, degeneracies can arise in three ways: (i) due to symmetries of the system; (ii) accidental degeneracies; and (iii) emergent degeneracies. For degeneracies due to exact symmetries of the bare Hamiltonian, the solution is to label states by their associated conserved quantum numbers and block-diagonalize the Hamiltonian separately in each quantum number subspace. A bi-orthonormal basis can then be defined separately in each block and different degenerate components of a symmetry multiplet are treated independently. For accidental degeneracies, often numerical error even at machine precision level is sufficient to distinguish states and eliminate problems with bi-orthonormalization. These issues were discussed in a different context for NRG calculations in Ref.~\cite{supp_lee2016adaptive}. We note that the procedure is stabilized by simply using 128-bit precision numerics, which is typically enough to distinguish accidental degeneracies, which are of course always approximate in practice. Another simple solution is to add to the Hamiltonian a physically inconsequential disorder perturbation of very small magnitude, which has the effect of lifting the degeneracies. Finally, in the context of quantum impurity problems, we note that low-energy fixed points can have larger \textit{emergent} symmetries than the bare model Hamiltonian. For example the one-channel, spin-$\tfrac{1}{2}$ anisotropic Kondo model has an \textit{isotropic} strong coupling stable fixed point~\cite{supp_Hewson1993}; whereas the two-channel Kondo model has a large emergent SO(8) symmetry at its critical point~\cite{supp_affleck1995conformal}. In these cases, one might expect additional degeneracies that cannot be separated into distinct quantum number blocks. However, these emergent symmetries only pertain asymptotically after very many NRG iterations, and at low energies. In practice, the degeneracies near the fixed point are always approximate and again, the use of high-precision numerics solves the problem. 
	
	%%%%%%%%%%%%%%%%%%%%
	
	\section{Iterative construction and diagonalization in NH-NRG}\label{app_IterativeConstruction}
	In the following we assume that left and right vectors of NH matrices are bi-orthonormal. 
	The NRG procedure is defined by the recursion relation,
	\begin{eqnarray}\label{eq:recursion}
		\hat{H}_{N+1}=\hat{H}_N + \hat{T}_{N+1}
	\end{eqnarray}
	which is initialized by $\hat{H}_0$, consisting of the impurity degrees of freedom and the Wilson chain `zero' orbital. Here the operator $\hat{T}_{N+1}=\hat{T}_{N+1}^a+ \hat{T}_{N+1}^b+\hat{T}_{N+1}^c$ is defined by,
	\begin{eqnarray}\label{eq:T}    \hat{T}_{N+1}^a=\epsilon_{N+1}^{\phantom{\dagger}}\sum_{\sigma} f_{N+1\sigma}^{\dagger} f_{N+1\sigma}^{\phantom{\dagger}} \qquad ;\qquad \hat{T}_{N+1}^b = t_N^{\phantom{\dagger}}\sum_{\sigma} f_{N\sigma}^{\dagger} f_{N+1\sigma}^{\phantom{\dagger}} \qquad ; \qquad \hat{T}_{N+1}^c =t_N^{\phantom{\dagger}} \sum_{\sigma} f_{N+1\sigma}^{\dagger} f_{N\sigma}^{\phantom{\dagger}}
	\end{eqnarray}
	At step $N+1$ of the iterative diagonalization process, we add on the new Wilson chain site $\ket{k}_{N+1}$, where the index $k=\{0,-1,+1,2\}$ labels the four possible configurations of that site, $\ket{k}_{N+1}=\{ \ket{-},\ket{\downarrow},\ket{\uparrow},\ket{\uparrow\downarrow} \}$ respectively. Since the part of the Hamiltonian describing the Wilson chain is Hermitian, the left and right eigenstates for the isolated Wilson orbital $\ket{k}_{N+1}$ are equal and so we do not specify a $L,R$ superscript. At this step we  need to construct the Hamiltonian matrix $\mathbf{H}^b_{N+1}$  with the following matrix elements,
	\begin{align}\label{eq:Hbelem}
		[\mathbf{H}^b_{N+1}]_{kr,k'r'} = \bra{N+1;k;r}_{b}^{L} \hat{H}_{N+1}^{\phantom{\dagger}} \ket{N+1;k';r'}_{b}^{R} 
	\end{align}
	where the $b$ subscript denotes that these are \textit{basis} states (rather than eigenstates), which are decomposed as,
	\begin{gather}\label{eq:bstate}
		\ket{N+1;k;r}_{b}^{L(R)} = \ket{k}_{N+1} \otimes \ket{N;r}_{d}^{L(R)} 
	\end{gather}
	for left(right) basis states. The latter are given in terms of the left(right) eigenstates in the \textit{diagonal} representation ($d$ subscript) of the previous iteration, denoted $\ket{N;r}_{d}^{L(R)}$. Therefore, these satisfy $\hat{H}_N \ket{N;r}_{d}^{R}=E_N(r)\ket{N;r}_{d}^{R}$ and $\bra{N;r}_{d}^{L}\hat{H}_N =\bra{N;r}_{d}^{L}E_N(r)$ where $E_N(r)$ are the complex eigenvalues of the previous iteration.
	
	We therefore have four terms to compute from Eqs.~\eqref{eq:recursion}, \eqref{eq:T}:
	\begin{align}
		&\bra{k}_{N+1} \bra{N;r}_{d}^{L} \hat{H}_N \ket{k'}_{N+1} \ket{N;r'}_{d}^{R} \;, \label{eq:HN}\\
		&\bra{k}_{N+1} \bra{N;r}_{d}^{L} \hat{T}_{N+1}^a \ket{k'}_{N+1} \ket{N;r'}_{d}^{R} \;, \label{eq:Ta}\\
		&\bra{k}_{N+1} \bra{N;r}_{d}^{L} \hat{T}_{N+1}^b \ket{k'}_{N+1} \ket{N;r'}_{d}^{R} \;, \label{eq:Tb}\\
		&\bra{k}_{N+1} \bra{N;r}_{d}^{L} \hat{T}_{N+1}^c \ket{k'}_{N+1} \ket{N;r'}_{d}^{R} \;. \label{eq:Tc}
	\end{align}
	
	%#################
	
	Since $\hat{H}_N$ comprises only even products of operators and does not act on degrees of freedom in orbital $N+1$, Eq.~\eqref{eq:HN} simplifies to:
	\begin{eqnarray}
		\innerprod{k}{k'}_{N+1} \bra{N;r}_{d}^{L} \hat{H}_N \ket{N,r'}_{d}^R = \delta_{kk'}\innerprod{N;r\phantom{'}}{N;r'}^{L,R}_{d} E_N(r') = \delta_{kk'}\delta_{rr'} E_N(r)\label{eq:HN2} \;.
	\end{eqnarray}
	
	Similarly, in Eq.~\eqref{eq:Ta} $\hat{T}_{N+1}^a$ consists of a number operator acting only on degrees of freedom of orbital $N+1$ and so reduces to,
	\begin{eqnarray}
		\bra{k}_{N+1} \hat{T}_{N+1}^a \ket{k'}_{N+1} \innerprod{N;r}{N;r'}^{L,R}_{d}  = \delta_{kk'}\delta_{rr'}\epsilon_{N+1}|k|  \label{eq:Ta2} \;,
	\end{eqnarray}
	where we used the fact that when using our convention for the index $k$, the spin-summed occupation number for state $\ket{k}_{N+1}$ in orbital $N+1$ is $n_k=|k|$.
	
	Eqs.~\eqref{eq:Tb} and \eqref{eq:Tc} are more complicated since they connect the part of the chain spanned by $\hat{H}_N$ to the added orbital $N+1$. To make progress we insert the identity,
	\begin{equation}
		\ID_{N+1} =\sum_{m,s} \ket{m}_{N+1}\ket{N;s}^{R}_{d}\bra{N;s}^{L}_{d}\bra{m}_{N+1} \;,
	\end{equation}
	between the creation and annihilation operators of $\hat{T}_{N+1}^b$ and $\hat{T}_{N+1}^c$ in Eq.~\eqref{eq:T}. Then Eq.~\eqref{eq:Tb} becomes,
	\begin{eqnarray}
		&t_N&\sum_{\sigma,m,s} \bra{N;r}_{d}^{L} \bra{k}_{N+1}  f_{N\sigma}^{\dagger} \ket{m}_{N+1}\ket{N;s}^{R}_{d}\bra{N;s}^{L}_{d}\bra{m}_{N+1} f_{N+1\sigma}^{\phantom{\dagger}} \ket{k'}_{N+1} \ket{N;r{'}}_{d}^{R} \\
		=&t_N&\sum_{\sigma,m,s} (-1)^k \innerprod{k}{m}_{N+1}
		\bra{N;r}_{d}^{L}f_{N\sigma}^{\dagger}\ket{N;s}^{R}_{d}\cdot \bra{m}_{N+1} f_{N+1\sigma}^{\phantom{\dagger}} \ket{k'}_{N+1} \cdot 
		\innerprod{N;s}{N;r'}_d^{L,R}  \\
		=&t_N&\sum_{\sigma} (-1)^k  M^\sigma_{k,k'} \cdot \bra{N;r}_{d}^{L}f_{N\sigma}^{\dagger}\ket{N;r'}^{R}_{d} \;,
	\end{eqnarray}
	where we have defined $M^{\sigma}_{k,k'}$ to denote the matrix element $\langle{k}|_{N+1}f_{N+1\sigma}|{k'}\rangle_{N+1}$, which is independent of the value of $N$. Note also that $(M^{\sigma}_{k,k'})^{\dagger} = M^{\sigma}_{k',k}$. The factor of $(-1)^k$ comes from the fermionic anticommutation when reordering operators. 
	
	Similarly for Eq.~\eqref{eq:Tc}, we obtain,
	\begin{eqnarray}
		t_N \sum_{\sigma} (-1)^{k'}  M^\sigma_{k',k} \cdot \bra{N;r}_{d}^{L}f_{N\sigma}\ket{N;r'}^{R}_{d} \;.
	\end{eqnarray}
	The nontrivial matrix elements $\bra{N;r}_{d}^{L}f_{N\sigma}^{\dagger}\ket{N;r'}^{R}_{d}$ and $\bra{N;r}_{d}^{L}f_{N\sigma}\ket{N;r'}^{R}_{d}$ must be computed at the previous step and saved. Note that they are not simple Hermitian conjugates of each other and must be calculated separately.\\
	
	From these expressions, one may construct the NH Hamiltonian $\hat{H}_{N+1}$ at step $N+1$ from information obtained at step $N$ -- specifically, the eigenvalues $E_N(r)$, and the matrix elements $\bra{N;r}_{d}^{L}f_{N\sigma}^{\dagger}\ket{N;r'}^{R}_{d}$ and $\bra{N;r}_{d}^{L}f_{N\sigma}\ket{N;r'}^{R}_{d}$. With $\mathbf{H}_{N+1}^b$ now constructed, we can diagonalize this matrix to obtain the eigenvalues $E_{N+1}$ and the left and right eigenvectors $\ket{N+1;r}_{d}^{L(R)}$. In particular, we can write $\mathbf{H}_{N+1}^b=\mathbf{U}_{N+1}^R \mathbf{H}_{N+1}^d (\mathbf{U}_{N+1}^L)^{\dagger}$ where  $\mathbf{H}_{N+1}^d$ is the \textit{diagonal} matrix of eigenvalues $E_{N+1}$ and  $\mathbf{U}_{N+1}^{R(L)}$ is a matrix whose columns are the right(left) eigenvectors. 
	This provides the set of complex eigenvalues $E_{N+1}$ needed for the next step.
	
	What about the matrix elements of the $f_{N+1\sigma}$ and $f_{N+1\sigma}^{\dagger}$ operators? These are also needed for the next step.
	To compute these, we expand the eigenstates as,
	\begin{gather}
		\ket{N+1;r}_{d}^{R(L)} = \sum_{m,s}U_{N+1}^{R(L)}(r;m,s)^{(\dagger)} \ket{N+1;m;s}_{b}^{R(L)} \nonumber\\
		\equiv \sum_{m,s}U_{N+1}^{R(L)}(r;m,s)^{(\dagger)} \ket{m}_{N+1}\ket{N;s}_{d}^{R(L)} \;.
	\end{gather}
	We use this to construct the  matrix element,
	\begin{eqnarray}
		\bra{N+1;r}_{d}^{L} f_{N+1\sigma}^\dagger \ket{N+1;r'}^R_d  &=& \sum_{\substack{m,s \\ m',s'}} U^{L}_{N+1}(r;m,s)^{\dagger}  U^{R}_{N+1}(r';m',s')  \innerprod{N;s}{N;s'}_{d}^{L,R} \bra{m}_{N+1}f_{N+1\sigma}^{\dagger}\ket{m'}_{N+1} \qquad \\
		&=&\sum_{m,m',s}  M^\sigma_{m'm} \:U^{L}_{N+1}(r;m,s)^{\dagger}  U^{R}_{N+1}(r';m',s)  
	\end{eqnarray}
	and similarly
	\begin{eqnarray}
		\bra{N+1;r}_{d}^{L} f_{N+1\sigma} \ket{N+1;r'}^R_d  &=& \sum_{m,m',s}  M^\sigma_{mm'} \:U^{L}_{N+1}(r;m,s)^{\dagger}  U^{R}_{N+1}(r';m',s) 
	\end{eqnarray}
	Thus, we have all of the ingredients to proceed to the next step. In this way, the entire chain can be built up orbital by orbital, starting from $\hat{H}_0$, which one explicitly constructs `by hand' in the initialization step.\\
	
	Without truncation, the Fock space would of course grow by a factor of four at each iteration. However, due to the exponentially-decaying Wilson chain parameters, we have a scale separation from iteration to iteration that motivates a truncation to just the $N_k$ lowest-lying states at each iteration, meaning that the computational complexity of the NH-NRG calculation scales \textit{linearly} with $N$ rather than exponentially. Of course, with complex eigenvalues $E_N$ at each step, there is a subtlety about what is meant by `lowest lying', and there are several truncation schemes that one could envision. The simplest, and the one that is closest to that employed in regular Hermitian NRG, is to truncate to the lowest $N_k$ eigenstates ordered by ${\rm Re}(E_N)$. This turns out to be the most numerically stable and accurate scheme, which we have confirmed reproduces correctly the exact results of exact diagonalization in the non-interacting limit. These issues are explored in more detail in the following sections.
	
	%--\FIGURE\-----------------------
	\begin{figure*}[t!]
		\centering  
		\includegraphics[width=\linewidth]{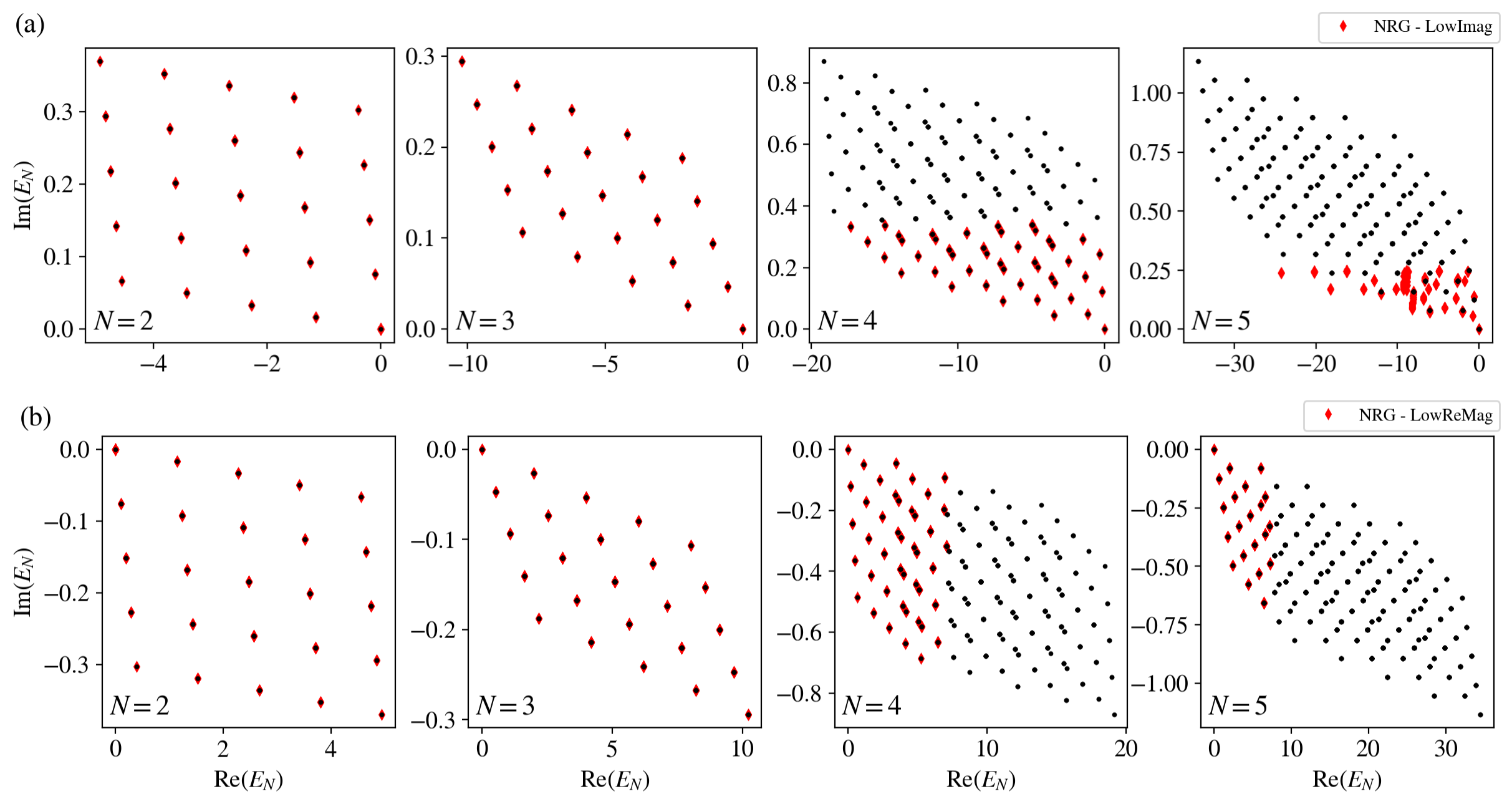}
		\caption{Illustration of alternate truncation schemes for NH-NRG on the non-interacting ($U_d=0$) NH AIM (also known as the non-Hermitian RLM). The plots are analogous to those in Fig.~E1 of the main text, and the same parameters are used. Top row panels (a) show truncation to the lowest $N_k$ states ordered by ${\rm Im}(E_N)$; bottom row panels (b) show a hybrid scheme in which the `ground state' with lowest ${\rm Re}(E_N)$ is first subtracted, and then states are sorted by magnitude, $|E_N-E_N^{gs}|$. NH-NRG results as red-diamonds, exact diagonalization results as black circle points.
		}
		\label{fig_rlm}
	\end{figure*}
	%--\FIGURE\--------------------- 
	
	%--\FIGURE\--------------------
	\begin{figure*}[t!]
		\centering 
		\includegraphics[width=\linewidth]{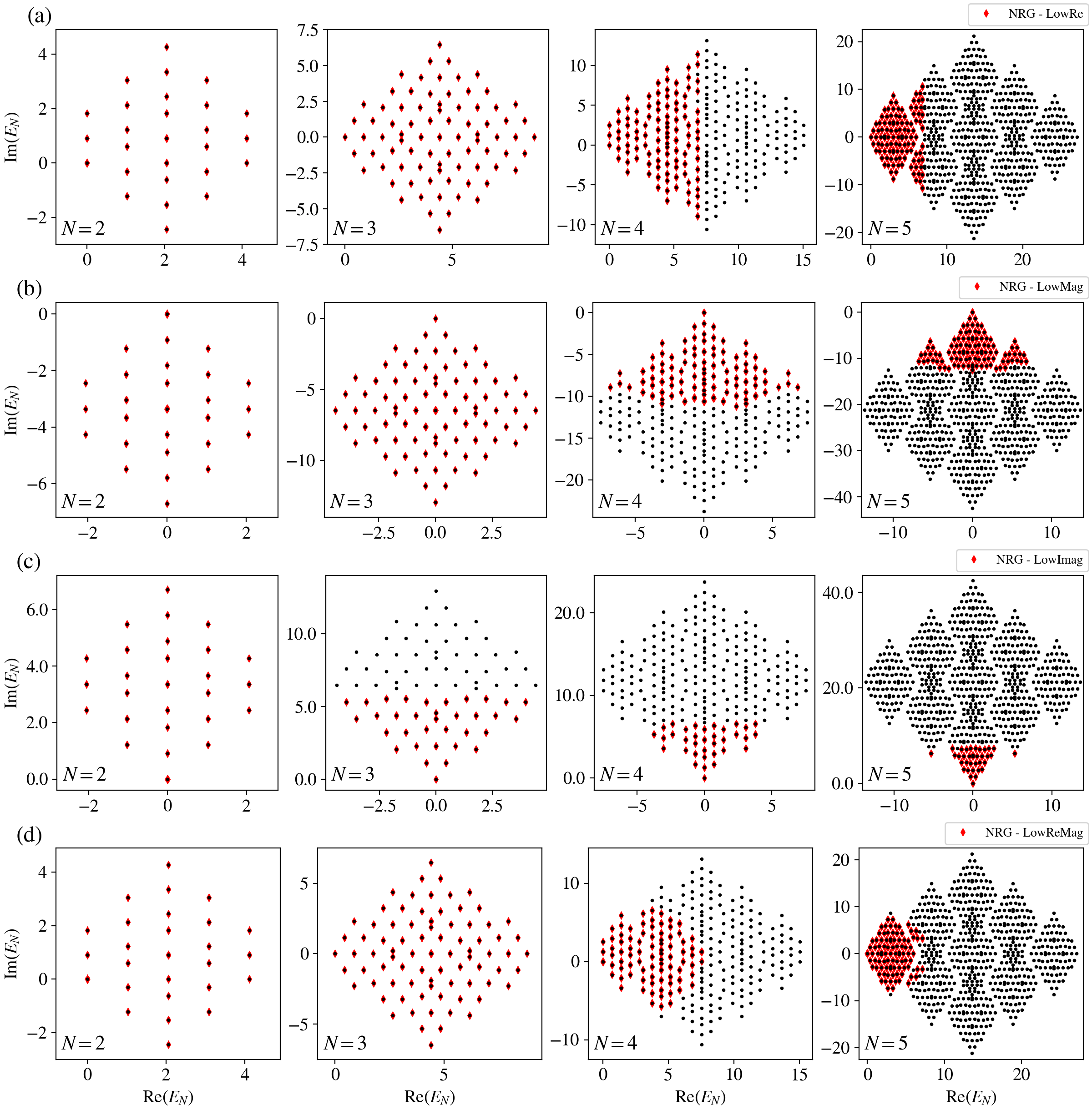} 
		\caption{Illustration different truncation schemes for NH-NRG for the free Wilson chain with imaginary on-site potentials. The four truncation schemes discussed in the text are shown, comparing NH-NRG results (red diamonds) with exact diagonalization (black circle points) for the complex eigenvalues of $\hat{H}_{N}$ for different iterations $N$. The full spectrum from exact diagonalization is shown in each case; NH-NRG reconstructs a different part of this spectrum due to the different truncation schemes employed. Plotted for $\Lambda=3$ and $N_k=400$.
		}
		\label{fig_wc}
	\end{figure*}
	%--\FIGURE\--------------------
	
	%%%%%%%%%%%%%%%%%%%%%%%%%%%%%%%%%%%%%%%%%%%%%%%%%%%%%%%%%%%%%%%%%%%%%%%%%%%%%%%
	\section{Alternative truncation schemes}\label{app_alt_truncs}
	
	\subsection{Non-Hermitian resonant level model}
	
	In the main text, we presented results for strongly-correlated quantum impurity problems obtained by NH-NRG using a truncation scheme (`LowRe') in which the lowest $N_k$ states were kept at each step, sorted by ${\rm Re}(E_N)$. In the \textit{End Matter} we presented some justification for that, by consideration of the non-interacting limit of the AIM ($U_d=0$), also known as the `resonant level model' (RLM). Being quadratic, the RLM can be solved exactly by diagonalizing the Hamiltonian in the single-particle sector (an $(N+2)\times (N+2)$ matrix at step $N$), and then constructing the $4^{N+2}$ many-particle states as simple product states from these -- a trivial combinatorial exercise. As such, the full eigenspectrum of the NH-NRG Hamiltonian $\hat{H}_N$ can be obtained by exact diagonalization for essentially any $N$ of interest, without any truncation, in this non-interacting limit. On the other hand, NH-NRG works in precisely the same way independently of $U_d$ and so the interacting AIM and non-interacting RLM are treated identically from an algorithmic point of view. The non-interacting RLM therefore provides a stringent check of our NH-NRG results. Fig.~E1(a) confirmed that truncation by lowest ${\rm Re}(E_N)$ correctly reproduces the exact eigenvalues at each step, for the retained states. One can also truncate by keeping the lowest $N_k$ states at each step, sorted by the absolute magnitude $|E_N|$, as shown in Fig.~E1(b) -- although in practice we found this to be less numerically stable. We dub this scheme `LowMag'.

	%--\FIGURE\-----------------------
	\begin{figure}[t!]
		\centering 
		\includegraphics[width=\linewidth]{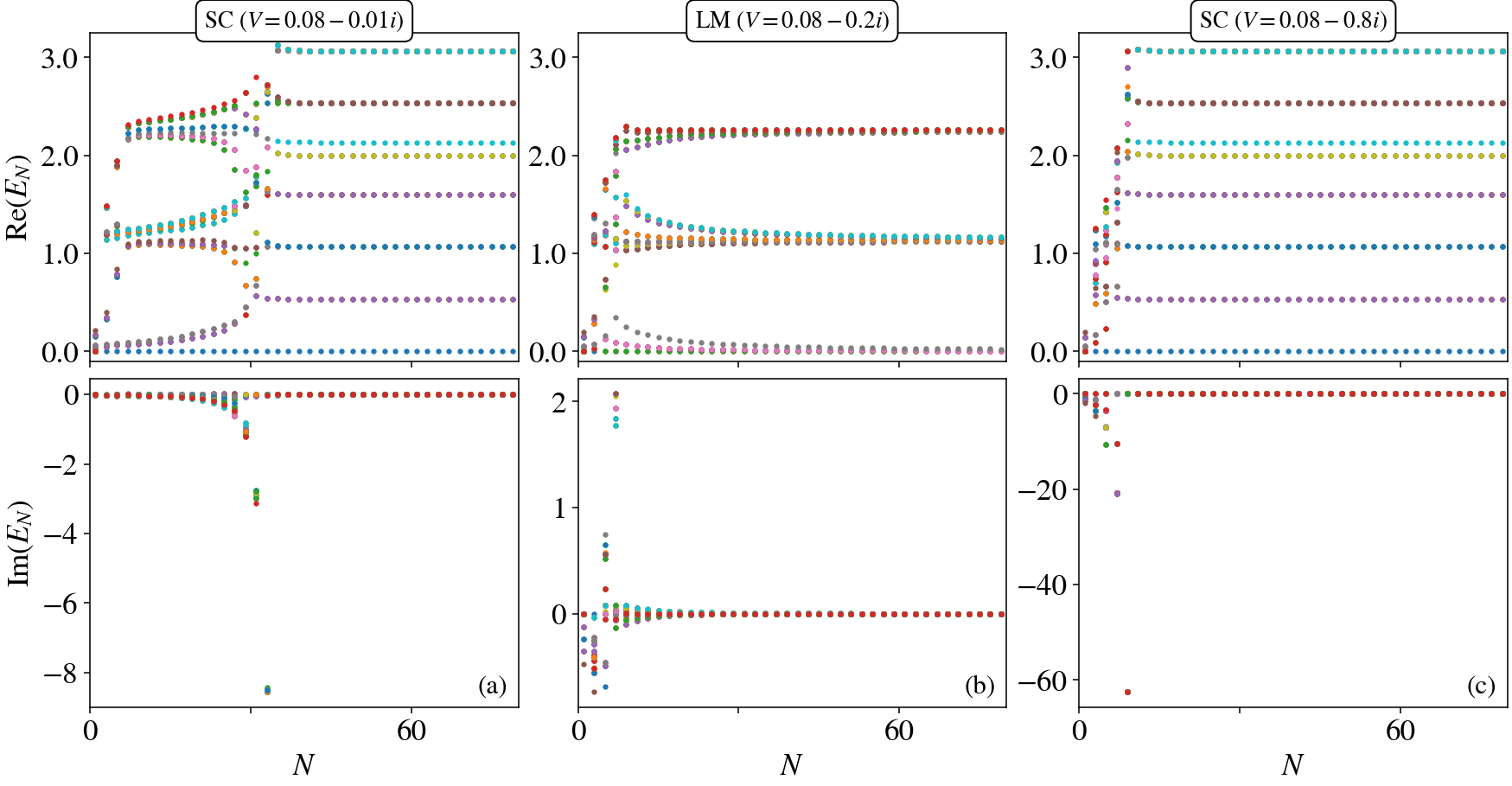}   
		\caption{Non-Hermitian Anderson impurity model ($U_d = 0.3$, $\epsilon_d=-0.15$, $\Lambda=3$, $N_k=400$, 128-bit precision): 
			Eigenvalue RG flow diagrams as ${\rm Im}(V)$ is made more negative, showing (a) SC phase; (b) LM phase; and (c) re-entrant SC phase.
		}
		\label{fig_AIM_3flow}
	\end{figure}
	%--\FIGURE\-----------------------
	
	In Fig.~\ref{fig_rlm} we consider two other truncation schemes. In the top row panels (a) we show truncation (`LowImag') to the lowest $N_k$ states ordered by ${\rm Im}(E_N)$, which targets a different set of kept states. While this method works initially, after a few steps it starts to break down. For $N=5$ we see that the NH-NRG eigenvalues no longer match those from exact diagonalization.
	
	In the bottom row panels Fig.~\ref{fig_rlm}(b), we use a hybrid scheme (`LowReMag') in which the `ground state' with the lowest ${\rm Re}(E_N)$ is first subtracted, and then states are ordered by their magnitude, $|E_N-E_N^{gs}|$. This truncation scheme also works very well and seems to be both accurate in reproducing the results of exact diagonalization, as well as being numerically stable. 
	
	In both cases we plot the \textit{rescaled} many-particle eigenvalues, comparing NH-NRG (red diamonds) with exact diagonalization (black circle points).

	%%%%%%%%%%%%%%%%%%%%%%%%%%%%%%%%%%%%%%%%%%%%%%%%%%%%%%%%%%%%%%%%%%%%%%%%%%%%%%%
	\subsection{Free Wilson chain with imaginary potentials}
	
	As a further demonstration, we consider NH-NRG for the free Wilson chain (no impurity). We introduce non-Hermiticity to the Wilson chain by using complex Wilson chain potentials. Specifically, we choose $\epsilon_n=-i t_n$, where $t_n$ are the usual Wilson chain hopping parameters for a metallic flat band with bandwidth $D=1$ as before. For the Hermitian symmetric flat-band Wilson chain, $\epsilon_n=0$, so introducing imaginary potentials down the chain simulates a kind of \textit{open} Wilson chain where each site is subject to dissipation and the states have a finite lifetime. This setup can be treated in NH-NRG very simply -- in practice we project out the impurity by setting $U_d=V=0$ and $\epsilon_d \gg D$. The resulting NH Wilson chain is simply a non-interacting tight-binding chain and can be solved exactly as per the results in the previous section.  In Fig.~\ref{fig_wc} we compare NH-NRG results (red diamonds) with those of exact diagonalization of the tight-binding model (black circle points), for the four truncation schemes discussed above. We again give results for the \textit{rescaled} many-particle eigenvalues. The results vividly show that NH-NRG works well in all cases, just reconstructing different parts of the spectrum when different truncation schemes are used.

	%--\FIGURE\-----------------------
	\begin{figure}[t!]
		\centering 
		\includegraphics[width=\linewidth]{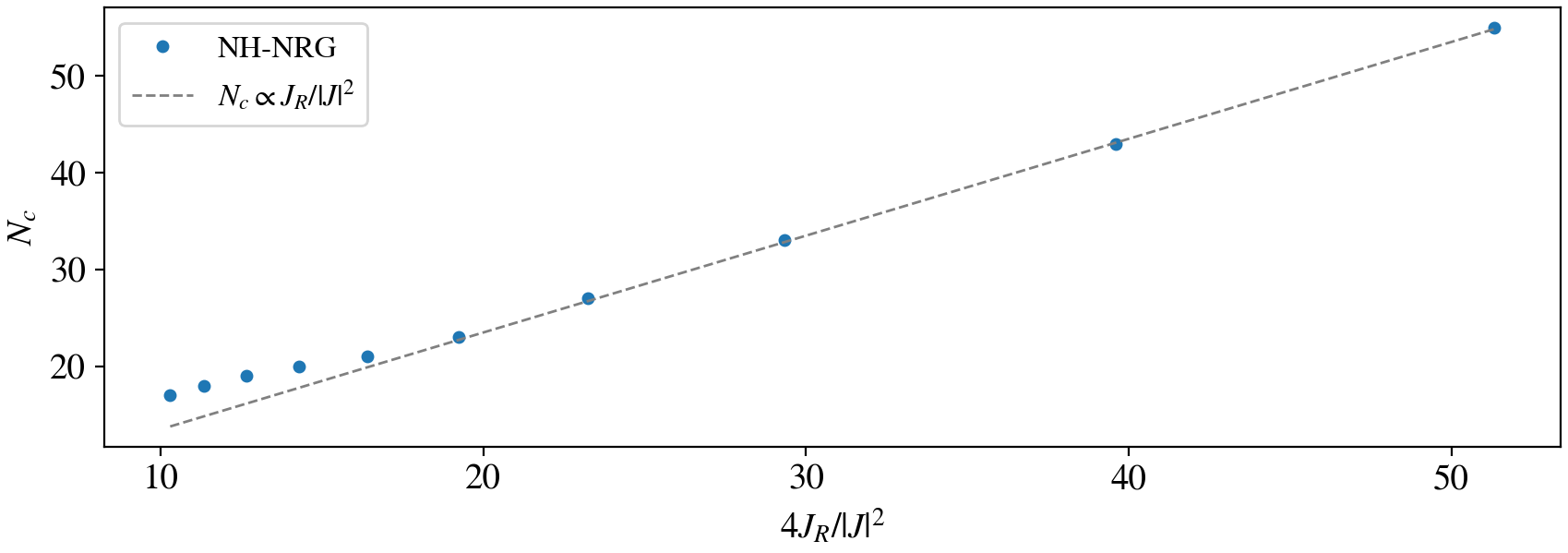}   
		\caption{Kondo temperature for the non-Hermitian Kondo model at weak coupling. The crossover iteration $N_c$ between LM and SC fixed points is extracted from NH-NRG eigenvalue flow diagrams for various $J_R$ and $J_I$ in the weak coupling (small $|J|$) regime. Shown for $\Lambda=3$, $N_k=400$. }
		\label{fig_Tk}
	\end{figure}
	%--\FIGURE\-----------------------
	
	%%%%%%%%%%%%%%%%%%%%%%%%%%%%%%%%%%%%%%%%%%%%%%%%%%%%%%%%%%%%%%%%%%%%%%%%%%%%%%% 
	
	%%%%%%%%%%%%%%%%%%%%%%%%%%%%%%%%%%%%%%%%%%%%%%%%%%%%%%%%%%%%%%%%%%%%%%%%%%%%%%%
	\section{Additional Anderson Impurity Model data}\label{app_AIM}
	In the main text we presented NH-NRG results for the NH AIM. Here in Fig.~\ref{fig_AIM_3flow} we show that by increasing the magnitude of the imaginary part of the impurity-bath hybridization $V$, one first observes a quantum phase transition from SC to LM, and then back to SC. This re-entrant Kondo behavior is predicted from the NH Kondo model (see Fig.~1(a) of the main text), but is also accessible in the parent AIM. 
	For strong enough ${\rm Re}(V)$ the LM phase disappears entirely. Thus the topology of the phase diagrams for Kondo and Anderson models is the same (albeit that naturally the details are somewhat different).
	This lends further support to the mapping between AIM and Kondo in the non-perturbative strong-coupling regime beyond Schrieffer-Wolff.
	
	We note that the Schrieffer-Wolff transformation between dissipative AIM and NH Kondo model derived in Ref.~\cite{supp_vanhoecke2024kondozenocrossoverdynamicsmonitored} gives strictly antiferromagnetic ${\rm Re}(J)>0$ and ${\rm Im}(J)<0$. This is the regime we focused on for the NH Kondo model in Fig.~1. With NH-NRG we indeed found that the ferromagnetic regime ${\rm Re}(J)<0$ was not accessible within the NH AIM. However, the ferromagnetic NH Kondo model might be interesting to study in its own right. We leave this to future work.
	
	%%%%%%%%%%%%%%%%%%%%%%%%%%%%%%%%%%%%%%%%%%%%%%%%%%%%%%%%%%%%%%%%%%%%%%%%%%%%%%%
	\section{Kondo Temperature}\label{app_Tk}
	
	Here, we numerically extract the crossover iteration number $N_c$, characterizing the flow between LM and SC fixed points from the RG flow diagrams of the NH-NRG. In Fig.~\ref{fig_Tk}, we plot the extracted $N_c$ as a function of the complex coupling $J = J_R - iJ_I$. At weak coupling (large $N_c$), we find excellent agreement with the predicted form of $T_K$ discussed in the main text. The running NRG energy scale~\cite{supp_wilson1975renormalization} is given by $E\sim D \Lambda^{-N/2}$ and so we identify the Kondo `temperature' $T_K\sim D\Lambda^{-N_c/2}$ in terms of the crossover iteration $N_c$. Our data is consistent with the relation $T_K\sim D e^{-2DJ_R/|J|^2}$ which implies $N_c=a+ bJ_R/|J|^2$ with $a$ an irrelevant constant that depends on the specific definition of $T_K$ used, and $b=4/\ln\Lambda$. The behavior of $T_K$ at stronger coupling was found to be more complicated.

	%%%%%%%%%%%%%%%%%%%%%%%%%%%%%%%%%%%%%%%%%%%%%%%%%%%%%%%%%%%%%%%%%%%%%%%%%%%%%%%
	\section{Critical point of the NH Kondo model}\label{app_crit}
	
\begin{figure}[t!]
		\centering 
		\includegraphics[width=\linewidth]{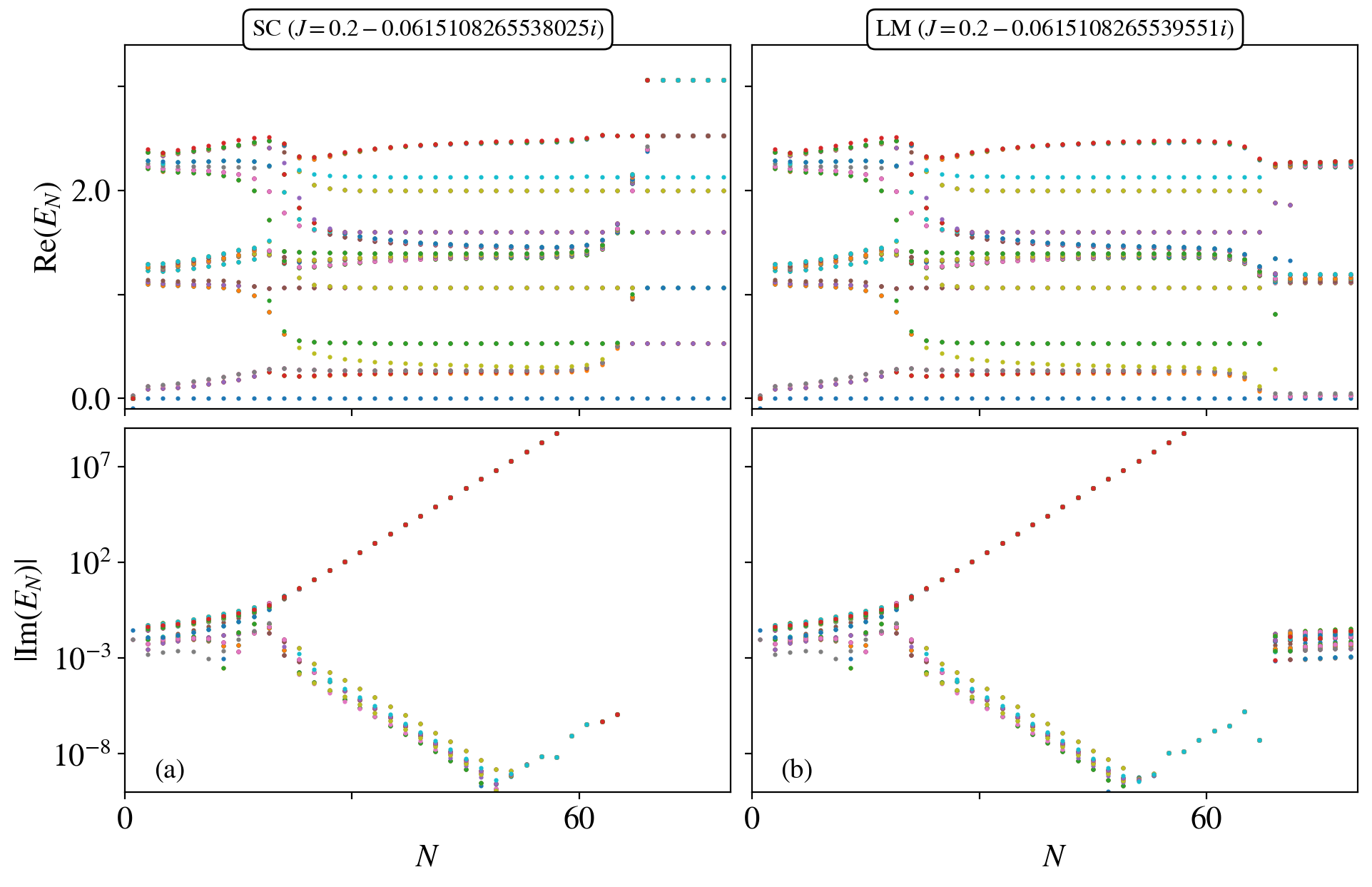}   
		\caption{
			RG flow of the NH-NRG complex eigenvalues $E_N$ with iteration number $N$, showing the real and imaginary parts in the top and bottom panels, for representative systems in the SC (a) and LM (b) phases.
			Shown for $\Lambda=3$, $N_k=200$ at 128-bit precision. }
		\label{fig_crit_RG}
	\end{figure}

	In the main text we identified a phase transition in the metallic NH Kondo model between SC and LM phases for $J_R>0$ as a function of $J_I$. In Fig.~\ref{fig_crit_RG} we show the RG flow on either side of the transition at $J=J_c$, obtained by NH-NRG. We tune $J_I$ very close to the transition in both cases, and see an extended RG flow in the vicinity of a novel critical fixed point. The critical fixed point is not of LM or SC type, and cannot be understood as a simple mixture of LM and SC states. In particular, we see a diverging imaginary part to the NRG energy levels at the critical point, with ${\rm Im}(E_N)$ growing exponentially with $N$. When the system eventually crosses over to either SC or LM, the imaginary part of the eigenvalues disappears, and the usual Hermitian Kondo fixed point structures emerge.  We therefore conclude that the transition is controlled by an unusual non-Hermitian critical fixed point. 
	
	%--\FIGURE\-----------------------
	%--\FIGURE\-----------------------
	%--\FIGURE\-----------------------
	\begin{figure}[t!]
		\centering 
		\includegraphics[width=\linewidth]{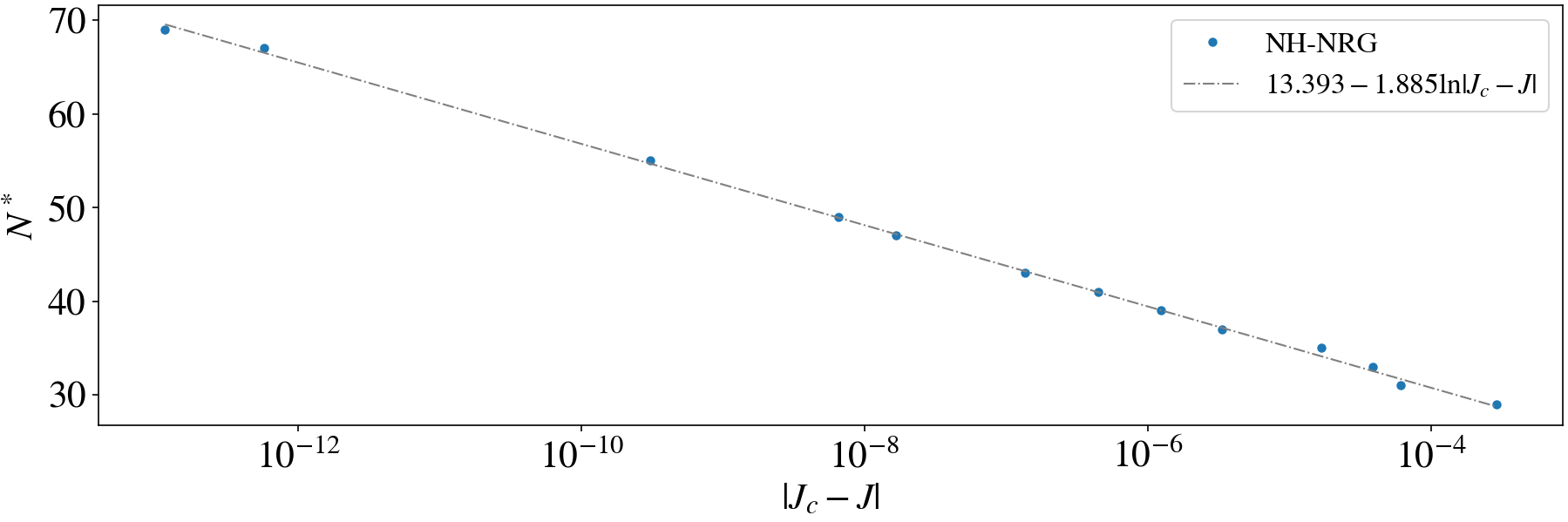}   
		\caption{Crossover scale for the non-Hermitian Kondo model near critical coupling. The crossover iteration $N^*$ between LM and SC fixed points is extracted from NH-NRG eigenvalue flow diagrams for various $J_R$ and $J_I$ in the regime of critical coupling $J_c$. Shown for $\Lambda=3$, $N_k=200$ at 128-bit precision. }
		\label{fig_crit_scale}
	\end{figure}
	%--\FIGURE\-----------------------
	
	In the vicinity of the transition near $J_c$, we identify a critical scale $T^*$ that vanishes as the transition is approached. From the NH-NRG data we identify a crossover iteration number $N^*$ for flow from the critical fixed point to either the LM or SC fixed points. In Fig.~\ref{fig_crit_scale} we show the evolution of $N^*$ with $|J-J_c|$, confirming the scaling behavior $N^*=a-b\ln|J-J_c|$. Since the corresponding crossover energy scale in NRG is $T^*\sim \Lambda^{-N^*/2}$, we may write $T^*\sim |J-J_c|^s$ with $s=\tfrac{1}{2}b\ln(\Lambda)$. With $\Lambda=3$ we extracted $b=1.885$ which yields $s\simeq 1$. Although for Hermitian systems this scaling might normally suggest a first-order transition, we note that the appearance of a distinct critical fixed point indicates otherwise. The non-Hermitian critical point appears to be rather exotic, possibly connected with an exceptional point of the model. A full understanding clearly requires further detailed study, and we leave this for future work.

	%%%%%%%%%%%%%%%%%%%%%%%%%%%%%%%%%%%%%%%%%%%%%%%%%%%%%%%%%%%%%%%%
	%\bibliography{Supp_Bib}
	
	%apsrev4-2.bst 2019-01-14 (MD) hand-edited version of apsrev4-1.bst
	%Control: key (0)
	%Control: author (8) initials jnrlst
	%Control: editor formatted (1) identically to author
	%Control: production of article title (0) allowed
	%Control: page (0) single
	%Control: year (1) truncated
	%Control: production of eprint (0) enabled
	%

%%%%%%%%%%%%%%%%%%%%%%%%%%%%%%%%%%%%%%%%%%%%%%%%%%%%%%%%%%%%%%%%%%%%%%%%%%%%%%%%%%%%%%%%%%

\end{document}